\documentclass{emulateapj}

\usepackage{natbib,graphicx,amsmath,amsthm,ulem,color}

\def\refnew#1{(\ref{#1})}
\def\be{\begin{equation}}
\def\ee{\end{equation}}
\newcommand{\beqn}{\begin{eqnarray}}
\newcommand{\eeqn}{\end{eqnarray}}
\newcommand{\bi}{\begin{itemize}}
\newcommand{\ei}{\end{itemize}}
\newcommand{\y}{}

\newcommand{\z}{}

\def\vhill{v_{\rm{H}}}

\def\erg{\, \rm erg}

\def\yrs{\, \rm yrs}
\def\s{\, \rm s}
\def\km{\, \rm km}
\def\cm{\, \rm cm}

\def\g{\rm g}

\def\deg{^\circ}

\begin{document} 
\title{Forming the Cold Classical Kuiper Belt in a light Disk}

\author{Andrew Shannon$^{1,3}$, Yanqin Wu$^{1}$ \& Yoram Lithwick$^{2}$}
\affil{$^1$Department of Astronomy and Astrophysics, University of Toronto, Toronto, ON M5S 3H4, Canada;}
\affil{$^2$Department of Physics and Astronomy, Northwestern University,
Evanston, IL 60208 and
Center for Interdisciplinary Exploration and Research in Astrophysics
(CIERA)}
\affil{$^3$Institute of Astronomy, University of Cambridge, Madingley Road, Cambridge CB3 0HA, UK}

\subjectheadings{Numerical simulations: Planetesimal growth}

\setcounter{equation}{0}
\setcounter{figure}{0}
\setcounter{table}{0}

\begin{abstract}
  Large Kuiper Belt Objects are conventionally thought to have formed
  out of a massive planetesimal belt that is a few thousand times its
  current mass.
  Such a picture, however, is incompatible with multiple lines of
    evidence.  Here, we present a new model for the conglomeration of
  Cold Classical Kuiper belt objects, out of a solid belt only a few
  times its current mass, or a few percent of the solid
  density in a Minimum Mass Solar Nebula.  This is made possible by
  depositing most of the primordial mass in grains of size centimetre
  or smaller. These grains collide frequently and maintain a
  dynamically cold belt out of which large bodies grow efficiently: an
  order-unity fraction of the solid mass can be converted into large
  bodies, in contrast to the $\sim 10^{-3}$ efficiency in conventional
  models.  Such a light belt may represent the true outer edge of
  the Solar system, and it may have effectively halted the outward
  migration of Neptune.  In addition to the high efficiency, our model can
also  produce a mass spectrum that peaks at an intermediate size, similar
  to the observed Cold Classicals, if one includes the effect of
  cratering collisions.  In particular, the observed power-law break
  observed at $\sim 30 \km$ for Cold Classicals, one that has been
  interpreted as a result of collisional erosion, may be primordial in
  origin.
\end{abstract}

\section{Introduction}
\label{sec:intro}

Beyond Neptune lies the Kuiper Belt, a population of remnant
planetesimals that were never incorporated into planets
(\citealp{1949MNRAS.109..600E,1993Natur.362..730J}; for a recent
review, see \citealp{2008ssbn.book.....B}).  Of particular interest
here is the Cold Classical Kuiper Belt, a population of
low-inclination ($i < 5\deg$), low eccentricity objects lying between
$42$ and $47$ AU
\citep{2001AJ....121.2804B,TrujilloBrown,2005AJ....129.1117E} with a
total mass of $\sim 0.1 M_{\oplus}$
\citep{2008AJ....136...83F,2009AJ....137...72F,2010A&A...520A..32V},\footnote{
  This estimate is uncertain by a factor of a few, affected by the
  assumed albedo, among other things.  }  and sizes up to $\sim 200
\km$ in radius \citep{2001AJ....121.1730L,2010Icar..210..944F}.

While the other Kuiper belt populations (the resonant and the
scattered bodies) appear to have been injected into the region via
interactions with planets
\citep{1993Natur.365..819M,1995AJ....110..420M,1997Natur.387..573L,2008ssbn.book..259G},
the quiescent orbits of the Cold Classical population cannot be
produced that way \citep{2008Icar..196..258L,2012ApJ...750...43D}, nor
would their delicate long-period binaries have survived such havoc
\citep{2010ApJ...722L.204P}.  The evidences strongly favour an {\it in
  situ} formation of the Cold Classical Kuiper Belt \citep[although
for a dissenting view, see][]{MorbidelliTrojan,2014Icar..232...81M}.
The color and size distributions of the Cold Classicals differ
markedly from the other populations
\citep{2000Natur.407..979T,2001AJ....121.1730L,2001AJ....121.2804B,2008Icar..194..758N,2009Icar..201..284B,2010Icar..210..944F,2014ApJ...782..100F,2010A&A...510A..53P,2011AAS...21730604F,2014ApJ...793L...2L},
further cementing its different origin.

The in-situ formation of the Cold Classicals, beyond the dynamical
ravage of the giant planets, allows us to place strong constraints on
their formation history. This gives hope that one can piece together
the giant puzzle, the origin of the outer solar system bodies, as well
as bodies in other outer planetary systems (manifested as extra-solar
debris disks), by scrutinizing this unique population.

\subsection{Formation in a heavy disk -- problems}
\label{subsec:exclude}

Previous studies for the formation of Kuiper belt bodies focussed on
their conglomeration out of a sea of kilometer-sized seeds, perhaps
instigated by the planetesimal formation scenario of
\citet{Safronov1969} and \citet{Goldreichward}.  Pairwise collisions
between these seeds produce large enough bodies that subsequently grow
by accreting the remaining seeds.  Meanwhile, the km-sized bodies are
viscously stirred by these growing big bodies.  The higher velocity
dispersion allows bigger bodies to grow more rapidly than smaller
bodies, quickly leaving them behind, in a process termed runaway
growth \citep{1978Icar...35....1G,WetherillStewart}. Eventually, the
runaway bodies stir the km bodies faster than they can accrete them
\citep{2008ApJS..179..451K,2010Icar..210..507O}.  Growth stalls.

In these models, growth lasts for a few collisional times of the
km-sized seeds, and it stalls when the biggest bodies have reached of
order Pluto size ($\sim 10^3 \km$). There is little evolution in
either the size spectrum or the total mass of large Kuiper belt
objects in the billions of years that follow, as the accretion time becomes longer than the system lifetime
\citep{1999ApJ...526..465K}.  In these models, the km-sized seeds
collide much less rapidly than they are stirred up. Collisional
cooling is unimportant for the dynamics, and we name this scenario
`collisionless growth'.

Two generic features are observed in all simulations of collisionless
growth
\citep{1978Icar...35....1G,WetherillStewart,Weidenschilling,1998AJ....115.2136K,2008ApJS..179..451K,2010Icar..210..507O,2011ApJ...728...68S}.
The efficiency of forming big bodies, defined as the fraction of
  the total initial mass that resides in bodies much larger than the
seeds, is generically $\sim 10^{-3}$ at the Kuiper belt distance
\citep{2011ApJ...728...68S,2015ApJ...801...15S}. The differential mass
distribution of these big bodies
\begin{equation} 
{\frac{dM}{d \log R}} \propto R^{n}\, ,
\label{eq:definedm}
\end{equation}
is flat for each logarithmic size decade, or $n=0$.\footnote{The more
  commonly used differential size distribution, $dN/dR \propto R^{-q}$
  is related as $q=4-n$.} These two features are simply understood
using the analytical arguments in \citet{2014ApJ...780...22L}:  they
arise because during conglomeration, the largest bodies regulate their own growth by exciting the dispersion of small bodies
to of order their own Hill velocity, or, small bodies remain
roughly `trans-hill' relative to the largest bodies at any
moment. Using our newly developed conglomeration code, we have
carefully tested this analytical understanding
\citep{2015ApJ...801...15S}.


For a formation efficiency of $\sim 10^{-3}$, a total mass in large
Cold Classical bodies of $\sim 0.1 M_{\oplus}$ will demand a
primordial solid mass of $\sim 100 M_{\oplus}$, comparable to the
solid mass expected for the Kuiper belt region in a Minimum Mass Solar
Nebula \citep[MMSN][]{1977Ap&SS..51..153W,Hayashi}.  However, such a
set-up appears to be excluded by multiple lines of evidence:
\begin{itemize}
\item Models for the migration of Neptune find that the primordial
  MMSN disk must end around 30 AU, or Neptune would have continued its
  outward march
  \citep{1984Icar...58..109F,1999Natur.402..635T,2004Icar..170..492G}.
\item If there had been $\gtrsim 1 M_{\oplus}$ of solids in km-sized
  bodies in the Kuiper Belt, long period Kuiper belt binaries would
  have been disrupted by impacts
  \citep{2012ApJ...744..139P,2011ApJ...743....1P}.
\item If the efficiency of forming the Cold Classical is low, $99.9\%$
  of the primordial mass (stored in small bodies like $1\km$
  planetesimals) has to be lost by collisional grind-down, followed by
  radiation pressure blow-out or Poynting-Robertson
  drag. However, attempts to model this
  loss process typically do not remove more than $\sim 90\%$ of the
  mass \citep{Kenyon:2004}, even in the most optimistic models
  \citep{PanSari}.

\item If one invokes dynamical ejection by giant planets to remove the
  primordial mass, disregarding the fact that the Cold Classicals are
  detached from the giant planets, since dynamical ejection has the
attribute of removing all body sizes equitably, this would have
required an even higher primordial mass. 

\end{itemize}
We conclude that the low formation efficiency, characteristic of
collisionless conglomeration, is not astrophysically viable for the
Cold Classicals.

The last line of argument also applies to the hot Kuiper belt objects
(the scattered and the resonant populations): if they were formed
with a similarly low efficiency, then Neptune, during its outward
migration, would have to carefully preserve these precious products,
or else the primordial disk has to be much more massive than the
corresponding MMSN value.  Lastly, old extra-solar debris disks,
comparable in age to the Solar System yet with dust luminosity three
or more orders of magnitude above that in the Kuiper belt
\citep{2006ApJ...636.1098B}, require $\sim 10 M_\oplus$ in the form
of large bodies \citep{2011ApJ...739...36S}. This is also incompatible
with the low efficiency of collisionless growth.

So the presence of Cold Classicals, and perhaps also the hot Kuiper
belt objects and extra-solar debris disks, call for a new formation
scenario.

\subsection{Formation in a light disk}
\label{subsec:intro2}

In this work, we explore a new model of formation, where the large
KBOs we observe today were formed in a proto-Kuiper belt that is light in mass and contained mostly centimeter-sized grains, along
with some kilometer-sized seed planetesimals.  This approach is
motivated by the \citet{GLS} proposal where they introduced small
grains to shorten the formation timescale for Neptune and Uranus.
Dynamic cooling by collisions among the centimeter sized grains
reduces their random velocities, and fundamentally alters the
dynamics.  The encounter velocities between bodies are, in this case,
dictated by the Keplerian shear (the so-called `shear-dominated'
regime), rather than by intrinsic velocity dispersion. In this regime,
runaway growth is avoided, and large bodies grow until a formation
efficiency of order unity is achieved.

The now much higher efficiency permits the {\it in situ} formation
  of the Cold Classicals, within a few million years. The primordial
  belt only needs to have a mass of $\sim 0.1 M_{\oplus}$, comparable
  to the mass in large KBOs today, and orders of magnitudes lower than
  the MMSN value. This circumvents the pitfalls in models of
  collisionless growth.  We call such a primordial disk the `Minimum
  Mass Kuiper Belt' (MMKB).  The size spectrum from such a
  `collisional growth' may be made compatible with the observed KBOs.

Are we justified in adopting such an initial condition, with
kilometre-sized rocks swimming in swarms of centimetre grains?

Dust sticking within the protoplanetary disk may have converted much
of the microscopic grains to larger particles \citep[see review
by][]{BlumWurm} . This growth, however, {\z may be sabotaged by
  collisional fragmentation when the cm-range is reached
  \citep{Brauer}, or arrested by bouncing, again at centimeter sizes
  \citep{2010A&A...513A..57Z}}. This motivates us to choose centimetre
for our small bodies.  {\z In addition, observations of comet
  103P/Hartley 2 show that much of its dusty coma is made up of grains
  $1-10 \cm$ in size \citep[][and its erratum]{2013Icar..222..634K},
  which may represent the material it formed from
  \citep{2015Icar..262....9K}. Lastly, centimeter-sized grains are
  observed to persist at tens of au for millions of years in
  protoplanetary disks
  \citep[e.g.,][]{2012ApJ...760L..17P,2012ApJ...747..136I,2013A&A...558A..64T}.
  This runs against the prevailing argument that aerodynamic drag,
  exerted by a gaseous disk on these grains, should bring them inward
  rapidly \citep{1977MNRAS.180...57W}, and points to missing
  ingredients in our understanding of their evolution.  }


A population of large seeds is necessary for our model.  The
gravitational potential of these bodies can accrete and retain small
bodies, so growth does not have to reply on particle sticking, which
may be difficult beyond the cm-size \citep{BlumWurm}.   These
  bodies should be of order kilometre or larger to allow rapid
  accretion in the initial stage, but their exact size and mass
  fraction do not impact our results (see \S \ref{subsec:dependence}).  A recent investigation of Comet 67P indicates that it is likely a
  result of gentle collision between two kilometre-sized bodies
  \citep{Massironi}. This suggests that such bodies may be common in
the Kuiper belt region during the early stage.
  Coincidentally, in a low-mass protoplanetary disk as the one assumed
  here, if the grains settle and concentrate toward the disk midplane
  and undergo gravitational instability \citep{Toomre,Goldreichward},
  the solid mass that lies within a Toomre wavelength corresponds to
  that of a kilometre body.  \footnote{\citet{Goldreichward,GLS} have
    argued that gravitational instability only form much smaller
    bodies as the angular momentum contained in a Toomre wavelength is
    too large to allow direct collapse into solid density. However,
    this is not seen in simulations of
    \citet{michikoshi07,michikoshi09}.}

  So in this work, we assume that by the time gas disk dissipates, one
  may be left with a host of km-sized planetesimals, surrounded by a
  sea of cm-grains. {\z These assumptions are critical to our results,
    but as is clear from the above discussion, while multiple pieces
    of circumstantial evidence suggest that such a set of initial
    conditions may be realistic, we are far from being able to prove
    it based on first-principle.}

In spirit, our conglomeration model is akin to the `pebble accretion'
model proposed for the formation of planetary cores in gaseous disks
\citep[e.g.][]{pebble,2015Icar..262....9K}. While in that model, gas
drag assists cm-sized `pebble' to be accreted onto large seeds
\citep{Rafikov,ormel}, pebbles here are cooled dynamically by mutual
collision, leading to efficient accretion onto large seeds.

\bigskip

This paper is arranged as follows. We present the theory and
simulations for collisionally-cooled growth in \S \ref{sec:cat} \& \S
\ref{sec:sim}, respectively. We then discuss the observational
implications of our results (\S \ref{sec:size}), including comparing
our predicted size spectrum against the observed one, and discuss
alternative formation models.

\section{Collisional Growth: Theory}
\label{sec:cat}

In the following subsections, we work out how collisionally-cooled
accretion proceeds, and derive the efficiency of formation.  The
  readers are referred to \citet{2014ApJ...780...22L} for a more
  general discussion that includes both collisional and collisionless
  conglomeration. Here, we assume that small body size is constant
  during the evolution.

\subsection{Prelude: Symbols and Values}
\label{sec:ic}

We adopt formulae for accretion, viscous stirring, collision, and
accretion as summarized by \citet{GLS}, along with their notation: big
bodies have radius $R$, velocity dispersion $v$, and surface density
$\Sigma$; small bodies have radius $s$, velocity dispersion $u$, and
surface density $\sigma$; all bodies have bulk density $\rho$ which we
take to be $\rho = 1.5 \g/\cm^3$, same as the Sun.  

We define an
important dimensionless quantity \be \alpha \equiv R/R_{\rm{H}} \sim
R_\odot/a \ , \ee which is the ratio of the physical size of a body to
its Hill sphere, or, approximately, the apparent angular size of the
Sun at distance $a$ ($R_\odot/a$, where $R_\odot$ is the solar radius).
For the Cold Classical region, $a\sim 45$ AU, so
$\alpha\sim 10^{-4}$. 

In the Kuiper belt, the orbital frequency is $\Omega\sim 0.02/$yr.
The Hill velocity for a body of size $R$ is, \be v_H\sim \Omega a
\left({R\over{ R_\odot}}\right) \sim 30 {\rm\ cm/s}\,
\left({R\over{100\km}}\right) \, \ .
\label{eq:vh} 
\ee The escape speed from the surface of such a body is $v_{\rm
  esc}\sim\alpha^{-1/2}v_H \sim 100 v_H$.

If we spread a mass of $0.1 M_\oplus$ uniformly within an annuli
  between 42 and 48AU, we obtain a surface density $\sigma =
  \sigma_{\rm MMKB}\approx 0.0016$~g/cm$^2$, where the subscript MMKB
  denotes `minimum mass Kuiper Belt'.

 In such a disk, collision time
between grains is short, \be t_{\rm col}\sim {\rho
  s\over\sigma\Omega}\sim 0.03 {\rm Myr} \, \left({s\over{1 \cm}}\right)
\, \left({{\sigma}\over{\sigma_{\rm MMKB}}}\right)^{-1} \, .
\label{eq:tcol}
\ee 

During the growth, we assume that the relevant velocity dispersions,
that of big bodies that dominate viscous stirring
(bodies with the highest $\Sigma(R)$), and that of small grains, are
both sub-hill relative to the stirrer.  \beqn v&<&v_H\, ,
\label{eq:vrange} \\
\alpha^{1/2}v_H<u&<&v_H \, .
\label{eq:range}
\eeqn So accretion is always in the sub-hill, orderly growth
regime. This ensures high efficiency of formation. We verify these
assumptions here.

During growth, velocity dispersion of small bodies reaches a
quasi-equilibrium, balancing collisional cooling among them and
viscous stirring by big bodies,
\be {1\over u}{du\over dt}\sim
{\Sigma\Omega\over\rho R}\alpha^{-2} {v_H\over
  u}-{\sigma\Omega\over\rho s} \approx 0 \ , \ee or, \beqn { u\over
  v_H} \sim {\Sigma\over\sigma}{s\over R}\alpha^{-2}\, .
\label{eq:uvh} 
\eeqn 
So the sub-hill condition ($u < v_H$) is satisfied if $s$ is
sufficiently small,
\begin{equation}
s \leq {\sigma\over{\Sigma}} R \alpha^2\, .
\label{eq:whererunawayends}
\end{equation}
We verify that this is satisfied throughout our simulation.

Velocity dispersion of the big bodies, on the other hand, is determined
by the balance between self-stirring and dynamical friction from small
bodies,
\be
{1 \over  v}{dv\over dt} \sim {\Omega\over\rho R}\alpha^{-2} \left(\Sigma{v_H\over v}-\sigma\right)\approx 0 \ .  
\ee 
Hence, 
\be
{v\over v_H}\sim {\Sigma\over\sigma} \, .
\label{eq:vvh}
\ee So until unity efficiency is reached, big bodies remain sub-hill,
consistent with inequality (\ref{eq:vrange}).

\subsection{Growth under Equal Accretion}
\label{sec:lt}

Under eq. \refnew{eq:range}, growth of
  big bodies by accreting small bodies proceeds at a rate
\citep{GLS},
\begin{equation}
\left. \frac{1}{R}\frac{dR}{dt}\right|_{\rm small} \sim \frac{\sigma \Omega}{\rho R}\alpha^{-1} \left(\frac{\vhill}{u}\right) \sim {\sigma^2\Omega\over\Sigma\rho s }\alpha
\label{eq:sacc}
\end{equation}

While accretion of small bodies dominates growth at early stages,
accretion of big bodies of comparable sizes becomes increasingly important over time.  We find that the growth
naturally enters into the so-called `equal accretion' phase where
the  two rates become comparable.

Big bodies grow by accreting one another at the rate 
\be 
\left.{1 \over R}{ dR\over dt}\right|_{\rm big}\sim {\Sigma\Omega\over\rho R}\alpha^{-3/2} \ .
\label{eq:bacc}
\ee Note that this rate is independent of $v$, whereas eq.
\refnew{eq:sacc} is $\propto 1/u$.  The rates differ because sub-Hill
big bodies lie in a flat disk, whereas small ones have an isotropic
velocity dispersion due to collisions
\citep{1992Icar...96..107I,Rafikov2003b,GLS}.

Defining the ratio between two modes of accretion to be
\be
f\equiv {d\ln R/dt\vert_{\rm small}\over d\ln R/dt\vert_{\rm big}} \sim  \left({\sigma\over\Sigma}\right)^2 \left({R\over s}\right) \alpha^{5/2}\, ,
\label{eq:fdef}
\ee we find that the system naturally tends toward $f \approx 1$. If
$f \gg 1$, as is the initial condition in our simulations, small body
dominates accretion and both $\Sigma$ and $R$ rise. However, the rise
of $\Sigma$ is faster since mass scales as $R^3$. This reduces $f$
toward unity. On the other hand, if $f \ll 1$ to start with, big
bodies grow by accreting each other. This increases $R$ while keeping
$\Sigma$ constant, thereby boosting $f$ toward unity. We conclude that
\be f\rightarrow 1
 \label{eq:ea} 
\ee 
at late times, a phase we call the `equal accretion' phase.  As a result, the fraction of mass in big bodies increases with $R$ as 
\be
{\Sigma\over\sigma}\, \sim \, \sqrt{R\over s}\alpha^{5/4}\, .
\label{eq:yssigma}
\ee This is one of the central results of this paper. The
  formation efficiency can reach unity for a sufficiently large
  $R/s$. We need small bodies to grow large bodies efficiently.

Equal accretion was first suggested by \cite{2011ApJ...728...68S}, but they considered the collisionless case only. In \citet{2015ApJ...801...15S}, we showed that in the collisionless case, equal accretion is a consequence of the dynamics, rather than the driving factor that it is in the collisional case.

Comparing eqs. \refnew{eq:whererunawayends} and \refnew{eq:yssigma},
it appears that the equal accretion phase can only commence after
\begin{equation}
R \geq s \alpha^{-3/2},
\label{eq:rlim}
\end{equation}
or $R\gtrsim 10\km (s/1\cm)$ for the Kuiper Belt.\footnote{Note that before equal accretion beings, $u \sim v_{H}$~\citep{2014ApJ...780...22L}}
 
Moreover, we verify that our assumption on small body dispersion,
en. \refnew{eq:range}, is satisfied during equal accretion. Combining
eqs. (\ref{eq:uvh}) and (\ref{eq:yssigma}), one obtains
\be {u\over v_H}\sim
\sqrt{s \alpha^{-3/2}\over R} \, .
\label{eq:uvh2} 
\ee Therefore, the ratio $u/v_H\sim 1$ when equal accretion
begins, and it falls to $u/v_H\sim \sqrt{\alpha}$ at its completion
($\Sigma \sim \sigma$).  The value of $u$ itself rises
with time (assuming $s$ is constant) as $u \propto R^{1/2}$.

\subsection{End of Equal Accretion}

If equal accretion proceeds indefinitely, eq.
(\ref{eq:yssigma}) implies that accretion reaches a radius
of completion (i.e. unity efficiency) at \be R\sim s\alpha^{-5/2}\ .
\label{eq:comp} 
\ee For the Cold Classical belt, this translates to $R \sim 100,000
\km (s/1\cm)$. In reality, however, equal accretion ends before
  that, when big bodies become oligarchs.  Spatial isolation of the
big bodies eliminates their mutual accretion and terminates the phase
of equal accretion. This occurs when the radial spacing between big
bodies ($\Delta a$) exceeds the size of their individual stirring and
feeding zones, $\sim R_H$ \citep{1998Icar..131..171K}.  Writing $2\pi
\Sigma a \Delta a=(4\pi/3)\rho R^3$, we find
\begin{equation}
{\Delta a\over R_H}\sim {2\over 3}{\rho\over\Sigma a}R^2\alpha\, ,
\label{eq:dah}
\end{equation}
Inserting eq. \refnew{eq:yssigma} into this and setting $\Delta a =
R_H$, we find the size at which oligarchy sets in is
\begin{eqnarray}
R_{\rm olig} & \sim & 
\left({{3 \sigma a}\over{2 \rho}}\right)^{2/3} \, \alpha^{1/6}\,
s^{-1/3} \nonumber \\
& \sim & 300 \km \left({{\sigma}\over{\sigma_{\rm MMKB}}}\right)^{2/3} 
\left({{s}\over{1\cm}}\right)^{-1/3}
\, ,
\label{eq:rolig}
\end{eqnarray}
where we have adopted $a=45AU$, $\alpha=10^{-4}$ and $\rho=1.5
\g/\cm^3$.

The efficiency of large body formation at this point is
(eq. \ref{eq:yssigma})
\begin{equation}
{{\Sigma}\over{\sigma}} \sim 5.5\% 
\left({{\sigma}\over{\sigma_{\rm MMKB}}}\right)^{1/3} \left({s\over{1\cm}}\right)^{-2/3}\, .
\label{eq:maxeff}
\end{equation}
And the timescale at this point is (eq. \ref{eq:sacc})
\begin{equation}
t_{\rm olig} \sim 16 
\left({{\sigma}\over{\sigma_{\rm MMKB}}}\right)^{1/3}
\left({s\over{1\cm}}\right)^{1/3}\, 
{\rm Myrs}\,\, ,
\label{eq:tolig}
\end{equation}

\subsection{End of Oligarchy}
\label{subsec:postolig}
Life continues during oligarchy.
Neighboring oligarchs continue to grow simultaneously (and
independently) by stirring and accreting their local small bodies
\citep{GLS}, at a rate that is prescribed by eq. \refnew{eq:sacc}. As
their spheres of influence begin to overlap, they scatter each other
to crossing orbits and merge.

Assuming small bodies do not reach high enough speed to fragment, and
that all bodies remain sub-hill (because $\Sigma < \sigma$ and $R \leq
s \alpha^2$, eqs. \ref{eq:uvh} \& \ref{eq:vvh}), unity
efficiency ($\Sigma \approx \sigma$) is reached at a time \beqn t_{\rm
  end} \sim {\rho s\over\sigma\Omega\alpha} \sim 300
{\left({s\over{1 \rm cm}}\right)}\,\ {\rm Myr} \, .
\label{eq:tgrow}
\eeqn 
and at a size
\begin{equation}
 R_{\rm iso}\sim \sqrt{\sigma a\over\rho\alpha}
\sim 800 \km  \left({{\sigma}\over{\sigma_{\rm MMKB}}}\right)^{1/2}
\, .  
\label{eq:riso}
\end{equation}
Such a body has the so-called isolation mass, where, moving on a
circular orbit, it sweeps up all material within its $R_H$ and
exhausts its own feed
\citep{1987Icar...69..249L,1998Icar..131..171K,GLS}.  The further
growth of isolation mass objects proceeds much more slowly, in
the timescale of orbital instabilities.

So the break-down of equal accretion, after the appearance of
  oligarchs, allows unity efficiency to be reached at a much smaller
  size than that in eq. \refnew{eq:rlim}.

More complicated processes, e.g., small body fragmentation, big body
long-range interactions, gap formation, may interfere with the above
picture.  We defer considerations of these dynamics to future work.

\section{Collisional Growth: Simulations}
\label{sec:sim}

\subsection{Numerical Code and Initial Conditions}
We simulate collisional accretion using a statistical particle-in-a-box
code, to compare against the theory presented in \S
\ref{sec:cat}.  Our code is described in \citet{2015ApJ...801...15S},
where we showed that various components in the code, including viscous
stirring, dynamical friction, accretion, collisional cooling, and
  collisional fragmentation, perform according to analytical
expectations. We also showed that the overall results from this code,
running under the collisionless condition, agree with results from
previous studies and conform to theoretical expectations.

In the current study, small grains collide frequently.  Depending on
the impact velocities, collisions can lead to cooling, cratering or
catastrophic destruction.  We adopt a zero coefficient of restitution
(very inelastic collision) for maximum cooling.
Catastrophic destruction occurs when the specific kinetic energy in
the impact, $0.5 [M_1 M_2/(M_1+M_2)^2] v^2$, with $v$ being the impact
velocity, exceeds the disruption threshold, which we adopt from
\citet{2009ApJ...691L.133S},
\begin{eqnarray} \nonumber
Q_{D}^* \approx 500 \left[ \left({{s_1}\over{1\cm}}\right)^3 
+ \left({{s_2}\over{1\cm}}\right)^3 
\right]^{-1/9} \left(v\over{1\cm/\s}\right)^{0.8} + \\ \nonumber
10^{-4} \left[ \left({{s_1}\over{1\cm}}\right)^3 
+ \left({{s_2}\over{1\cm}}\right)^3 
\right]^{0.4} \left(v\over{1\cm/\s}\right)^{0.8} \, \erg/\g, \\ \, 
\label{eq:Qstar}
\end{eqnarray}
where $s_1, s_2$, $M_1, M_2$ are the sizes and masses of the two
colliding particles, respectively.  This scaling applies for small
aggregates that are strength dominated.  For an equal-mass impact,
this threshold corresponds to an eccentricity of $e \sim 10^{-3}$ at
$s=1\cm$.  Similarly, an $1\km$ body should only be destroyed by
another $1\km$ body at an eccentricity $\sim 10^{-3}$, as they are
bound by self-gravity.  Speeds this high is not reached until near the
end of our simulations. As a result, catastrophic destruction is not a
significant process.

On the other hand, cratering (chipping away the target by a low-energy
impact) can be important. In the following, we first ignore cratering
but return to discuss its effects in \S \ref{subsec:cratering}.

Similar to \citet{2015ApJ...801...15S}, we trace the mass evolution
using logarithmic mass bins that are spaced by 0.1 dex. Each bin is
characterized by one eccentricity value. The reader is referred to
that paper for details on how big bodies are promoted in size, and for
our procedure to emulate isolation once oligarchs appear. For
simplicity, we consider only a single radial zone, tracking the size
spectrum and eccentricity as they evolve in time.  Random velocities
are assumed to be isotropic ($i \approx e$), except for big bodies for
which we assume $i \ll e$, i.e., they lie in a thin disk \citep{GLS}.

To simulate the isolation effect of oligarchy, we do not allow the large bodies to accrete each other, once they become oligarchs.

The initial surface density is taken to be $\sigma = \sigma_{\rm MMKB}
\approx 0.0016$~g/cm$^2$. We deposit $99.9\%$ in $s=1\cm$ bodies, and
the remaining $0.1\%$ in 1-km bodies.  {\z These choices are somewhat
  arbitrary. Fortunately, our results are little affected if mass in the large
  seeds is altered by two orders of magnitude up or down.}


All bodies are started with an initial eccentricity of $e = 10^{-6}$,
the Hill velocity of a $1-\km$ body. This is motivated by the argument
that, even when one starts from a different condition, a trans-hill
state is quickly achieved \citep{2014ApJ...780...22L}. This value is
also motivated by the fact that the particle disk we consider will be
Toomre unstable if the velocity dispersion falls below $e \sim
10^{-6}$.  As in \citet{2015ApJ...801...15S}, eccentricity evolution
is much faster than mass evolution, and the system relaxes before
significant mass evolution takes place.

\subsection{Results}

\begin{figure*}
  \centering
  {\includegraphics[width=0.49\textwidth,trim = 15 0 12 8, clip]{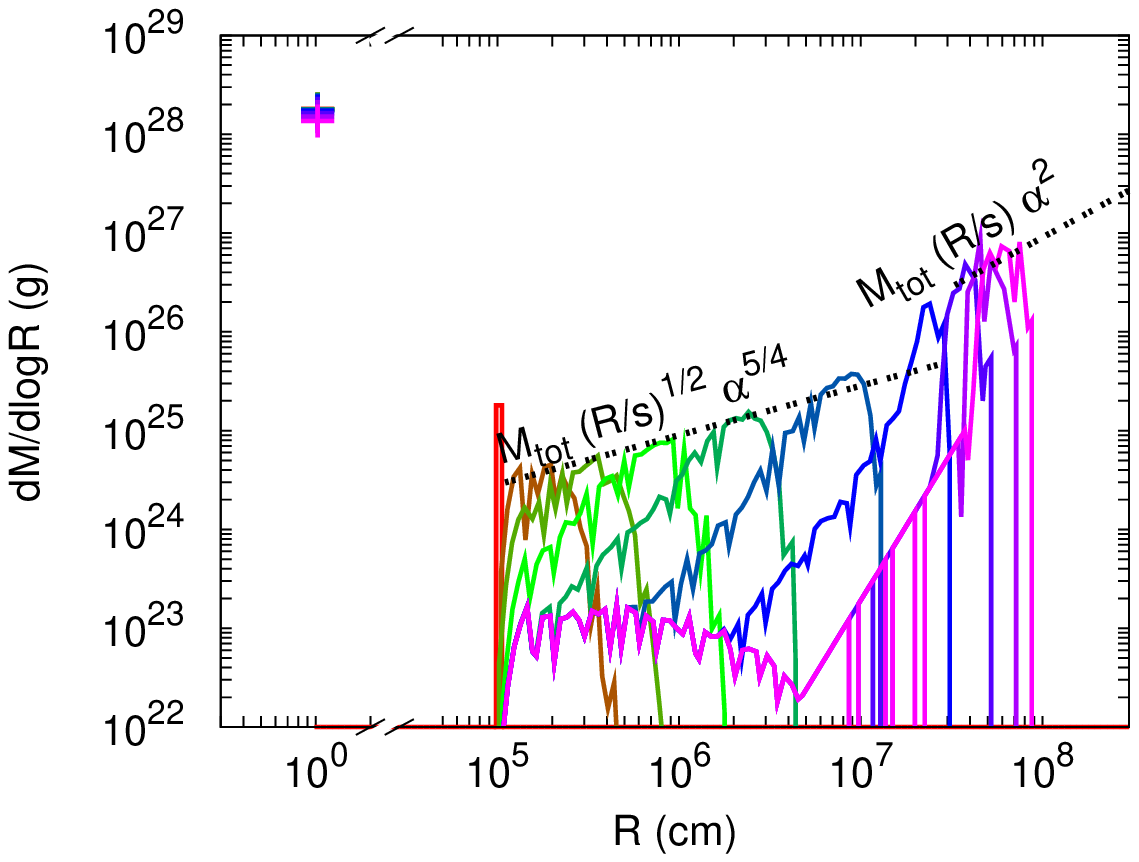}}
  {\includegraphics[width=0.49\textwidth,trim = 15 0 15 12, clip]{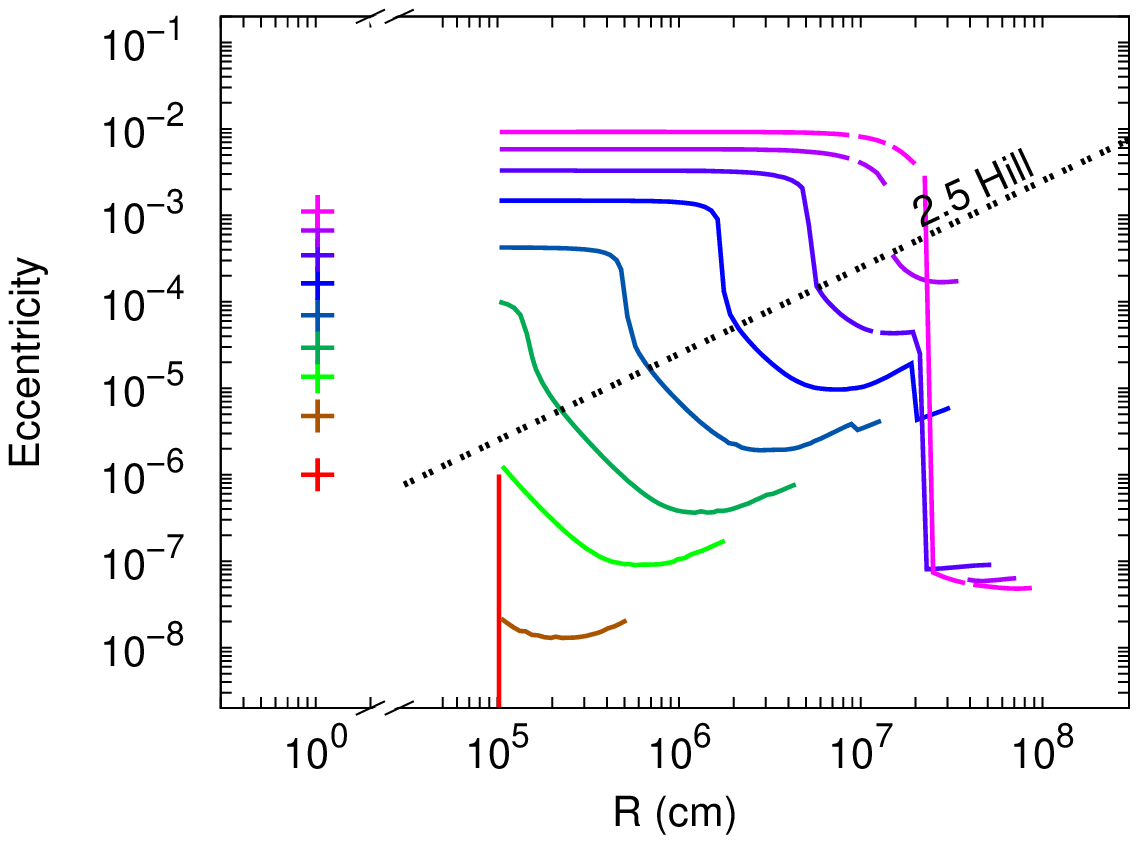}}
  \caption{ Collisional conglomeration in a low-mass disk. The left
    panel depicts the evolution of the differential mass spectrum, and
    the right that of eccentricity, both plotted as functions of
    particle size. Initially, $99.9\%$ of the mass is in cm-grains and
    $0.1\%$ in km-sized rocks.  Snapshots of the evolution are taken
    at $0$ (red),$ 1, 2, 4, 8, 16, 32, 64$ Myrs, and at $10$ Gyrs
    (magenta line).  The lower dashed line shows the predicted mass in
    the largest bodies as a function of $R$ (eq.  [\ref{eq:yssigma}]),
    valid during equal accretion.  At this stage, velocities of both
    the small and the largest bodies remain subhill (of the largest
    bodies).  Beyond $16$ Myrs, the big bodies become oligarchs and
    they grow by accreting only small bodies (but not each other), in
    the trans-hill regime (the upper dashed line). {\z There is no
collisional cratering and fragmentation included here.}}
\label{fig:plot318}
\end{figure*}

\begin{figure}
  \centering
  {\includegraphics[width=0.49\textwidth,trim = 0 0 0 0, clip]{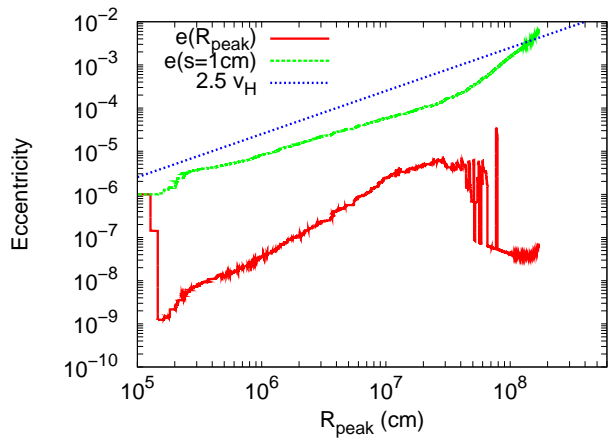}}
  \caption{Eccentricities for both small grains (green line) and
    $R_{\rm peak}$ bodies (red line), as functions of $R_{\rm peak}$,
    for the simulation presented in Fig. \ref{fig:plot318}.  $R_{\rm
      peak}$ is formally the size of large $\left(\geq 1
        \rm{km}\right)$~bodies carrying the most mass, and is roughly
    the size of the largest bodies at any given time. Before $R_{\rm
      peak}$ reaches $\sim 300\km$, the evolution is dominated by
    equal accretion with big and small bodies both remaining sub-hill
    (compare the blue-dotted and green lines, eq.
    \ref{eq:uvh2}). Once oligarchy sets in at $R_{\rm peak} \sim 300
    \km$, big-bodies are effectively isolated from each other's
    stirring and from accreting each other. Their eccentricities drop
    precipitously and they grow by accreting small bodies only. A
    trans-hill growth ensues.
}
\label{fig:plotST2}
\end{figure}

\begin{figure}
\begin{center}
  \includegraphics[width=.49\textwidth, trim = 0 0 -5 0, clip,
  angle=0]{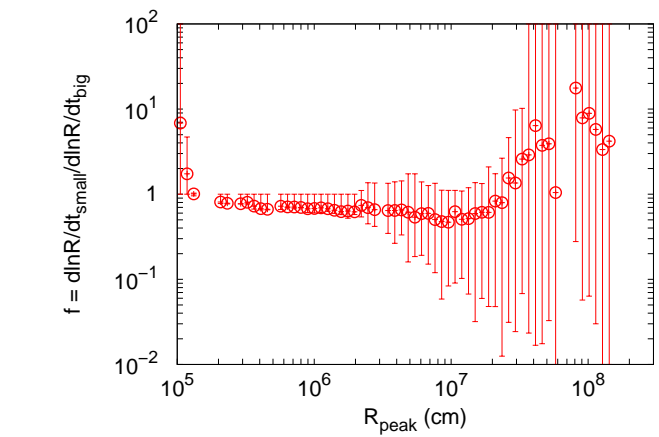}
  \caption{Here, we demonstrate equal accretion by plotting the ratio
    of two accretion rates for bodies of size $R_{\rm peak}$, that of
small body and that of $R\sim R_{\rm peak}$ bodies.
Points are the average ratio 
at each size, with error bars indicating the highest and the lowest
instantaneous values.  The system quickly finds the $f \sim
1$~equilibrium (eq. \ref{eq:ea}).  Growth proceeds in equal
accretion mode until $R_{\rm peak}$ reaches $\sim 300 \rm{km}$.}
\label{fig:equalaccretion}
\end{center}
\end{figure}

Results of the collisional growth are plotted in Figure
\ref{fig:plot318}, with Figs. \ref{fig:plotST2} and
\ref{fig:equalaccretion} providing diagnostic details.  Overall, the
simulation results confirm our analytical picture in \S
\ref{sec:cat} and we give a brief re-cap below.

Frequent collisions between small grains keep them cold and allow
efficient accretion onto km-sized seeds, which are also kept cool by
dynamical friction from the small grains.  Growth is rapid and within
$\sim 10^5 \yrs$, most seeds have doubled their sizes. The rise in big
body population accelerates cannibalism among them and when the big
bodies reach sizes $\sim 2\km$, smaller than that estimated in eq.
\refnew{eq:rlim}, we see that equal accretion has set in
(Fig. \ref{fig:equalaccretion}). 

Defining $R_{\rm peak}$ to be the size of bodies that dominate
stirring (also roughly the largest bodies at any time), we see that as
$R_{\rm peak}$ rises, both the largest bodies and the small grains
remain cold, $u, v \leq v_H (R=R_{\rm peak})$
(Fig. \ref{fig:plotST2}), in accordance with eqs. \refnew{eq:vvh} \&
\refnew{eq:uvh2}.  Equal accretion predicts that mass in $R \approx
R_{\rm peak}$ rises as $R_{\rm peak}^{1/2}$ (eq. \ref{eq:yssigma}), and
is indeed observed in the left panel of Fig. \ref{fig:plot318}. By 30
Myrs (eq. \ref{eq:tolig}), oligarchy is reached with $R_{\rm peak}
\sim R_{\rm olig} \sim 300 \rm{km} $, and the efficiency of
conglomeration at this point is $\sim 5\%$, as is estimated in
eq. \refnew{eq:maxeff}.  

At this stage, intermediate size bodies are severely depleted, caused
by both cannibalism among equal-sized bodies and by accretion onto
larger bodies. The resultant differential mass spectrum is
$d\Sigma/d\ln R \propto R^2$, or a differential number distribution of
$dN/d R \propto R^{-2}$. This is much more top-heavy compared to the
collisionless case where $dN/d R \propto R^{-4}$, and also more so
than that obtains assuming a trans-hill
collisional growth, $dN/dR \propto R^{-3}$,
\citep{2014ApJ...780...22L}. 

Continuing integration shows that near unity more precisely,
  $\left(\approx 25\%\right)$~efficiency is reached when $R \sim 800
\km$ (eq. \ref{eq:riso}),occurring after $\sim 300$ Myrs of evolution
(eq. \ref{eq:tgrow}).

\subsection{Collisional vs. Collisionless}

\begin{figure*}
  \centering
  {\includegraphics[width=0.48\textwidth,trim = 10 0 15 5, clip]{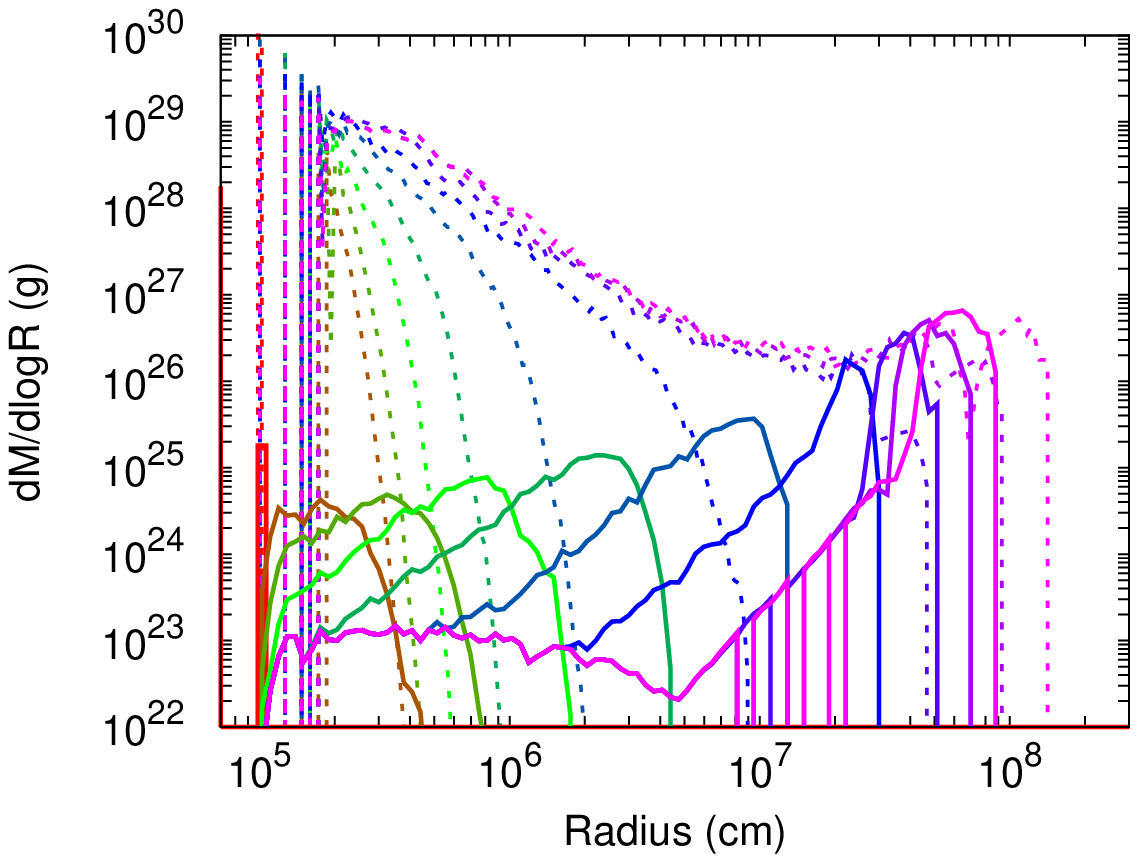}}
  {\includegraphics[width=0.50\textwidth,trim = 0 0 0 5, clip]{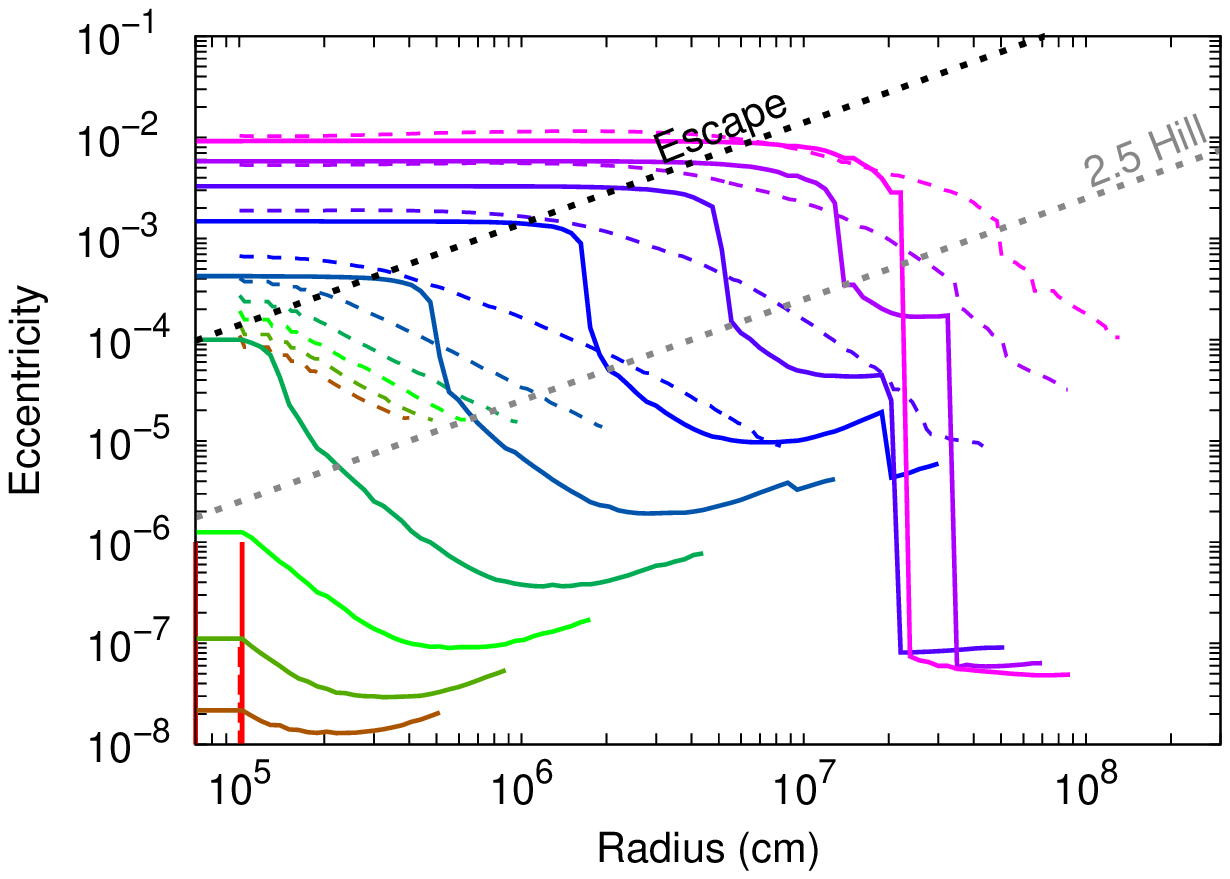}}
  \caption{A
    comparison between collisional (solid curves, not showing small
    grains, Fig. \ref{fig:plot318} here) and collisionless (dashed
    curves, Fig. 4 in \citet{2015ApJ...801...15S}) models of
    conglomeration. The former case
    begins with only $1\%$ mass of the latter. Snapshots for both
    cases are taken at $0, 1,
    2, 4, 8, 16, 32, 64, 128$ and $256$ Myrs.
    The efficiency of large body formation, measured at late times, is
    of order unity for the collisional case, and much lower ($\sim
    10^{-3}$) for the collisionless one, leading to similar numbers of
    large bodies being produced in both runs. The mass distribution is
    top-heavy for the collisional one, while flat for the
    collisionless one. 
  }
\label{fig:comparison2}
\end{figure*}

The dividing line between collisional and collisionless regimes lies
at $s/R \sim \alpha$ \citep{2014ApJ...780...22L}. Throughout our
simulations here, with $s/R \sim 10^{-5} (s/{1\cm})(R/{1\km})^{-1}
\leq 10^{-5}$, collisional cooling remains important, in contrast to our
study in \citet{2015ApJ...801...15S} where $s \sim 1\km$.  The 
different outcomes from these two sets of simulations are summarized
in Fig. \ref{fig:comparison2}, while readers are referred to
\citet{2014ApJ...780...22L} for an analytical understanding.

The collisionless growth has an efficiency of $10 \alpha \sim
10^{-3}$, while the collisional growth reaches an efficiency of $5\%$
even at $R = R_{\rm olig}$.  So to produce the same number of large
bodies, the former model requires a disk that is $\sim 100$ times more
massive than the collisional one. Moreover, the resultant size spectra
differs between the two cases. While the collisionless case deposits
equal mass per logarithmic decade, or $dN/dR \propto R^{-4}$, the
collisional case leads to a heavy depletion in intermediate sizes, with
most of the mass concentrated in the largest bodies, $dN/dR \propto
R^{-2}$.

Despite the substantially lower surface density, the collisional case
can produce the same $R_{\rm peak}$ at roughly the same timescale as
the collisionless one -- because the small bodies' speeds remain
  small, gravitational focusing aids accretion. Oligarchy,
at $R_{\rm peak} \sim 300\km$, is reached in both cases at about 30-60
Myrs.

\subsection{Cratering}
\label{subsec:cratering}

For the material strength law that we adopted
\citep{2009ApJ...691L.133S}, collisions cause catastrophic destruction
between equal-sized bodies at eccentricities $e \gtrsim 10^{-3}$ {\z
  (for both 1-cm and 1-km bodies; it is smaller for sizes in between)}, a
value reached only near the end of our simulation (see
Fig. \ref{fig:plot318}).  In contrast, cratering, where a fraction of
the target mass is removed, can occur at lower velocities and be
potentially important.  We study the impact of this process here.

We adopt the approach of \citet{2010Icar..206..735K} to write the
total ejected mass after a collision as
\begin{equation}
 m_e = \frac{\phi}{1+\phi}m\, ,
\end{equation}
where $m$ is the target mass, and $\phi$ is the ratio between the
specific impact energy to the critical specific energy for
catastrophic disruption, $Q^*_{D}$.  Moreover, {\z we write} the mass
of the largest fragment {\z as}
\begin{equation}
 m_f = \frac{\phi}{\left(1+\phi\right)^2} m\, .
\end{equation}
We deposit the total ejected mass ($m_e$) as {\z a power law in size
  between those with masses $m_f$ and 0, with $dN/dR \propto R^{-q}$
  and $q = 3.5$ \citep{Dohnanyi:1969}. So most of the fragment mass
  lies near $m_f$.  We place mass for bodies larger than $1 \mu m$ in
  the appropriate mass bin and follow them in the numerical
  integration, while bodies} below $1 \mu m$~are instantaneously removed
from the simulation, {\z as would be expected due to} radiation
pressure from the Sun. 

{\z Cratering allows a net mass-loss from the system. While the
  catastrophic fragmentation of a $1\mu m$ grain requires $e \geq
  0.01$ (a value barely reachable in our simulation, see
  Fig. \ref{fig:plot318}), cratering can remove mass at lower velocities.}

For the threshold of catastrophic destruction ($Q_D^*$), we adopt the
strength formula for small bodies, eq. \refnew{eq:Qstar},
{\z but} weaken it by a factor of $4$, while keeping the threshold for
bodies that are gravity-dominated (km-sized and above) unchanged. This
factor of $4$ is introduced to achieve a good match to observations in
\textsection \ref{subsec:large}.  We discuss varying this value in \S
\ref{subsec:dependence}. 

Moreover, to compensate for the fact that much of the initial mass is
ground down and blown out, 
we increase the initial surface density to $\sigma = 3\times
\sigma_{\rm MMKB}$, or $0.3 M_\oplus$ in the annulus from $42$ to $48$
AU. This elevated total mass still evades the constraint set by the
survival of binary KBOs and the migration stalling of Neptune.

\begin{figure*}
  \centering
  {\includegraphics[width=0.49\textwidth,trim = 15 0 15 0, clip]{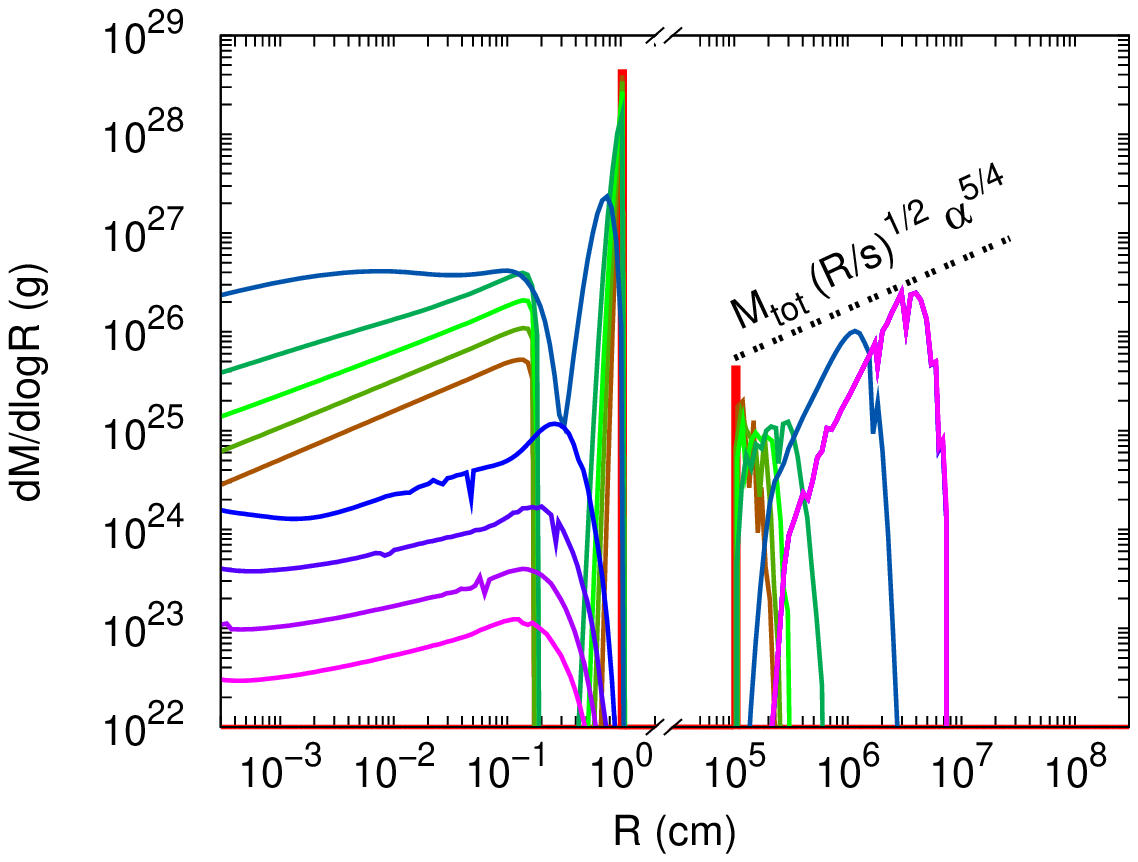}}
  {\includegraphics[width=0.49\textwidth,trim = 15 0 8 14, clip]{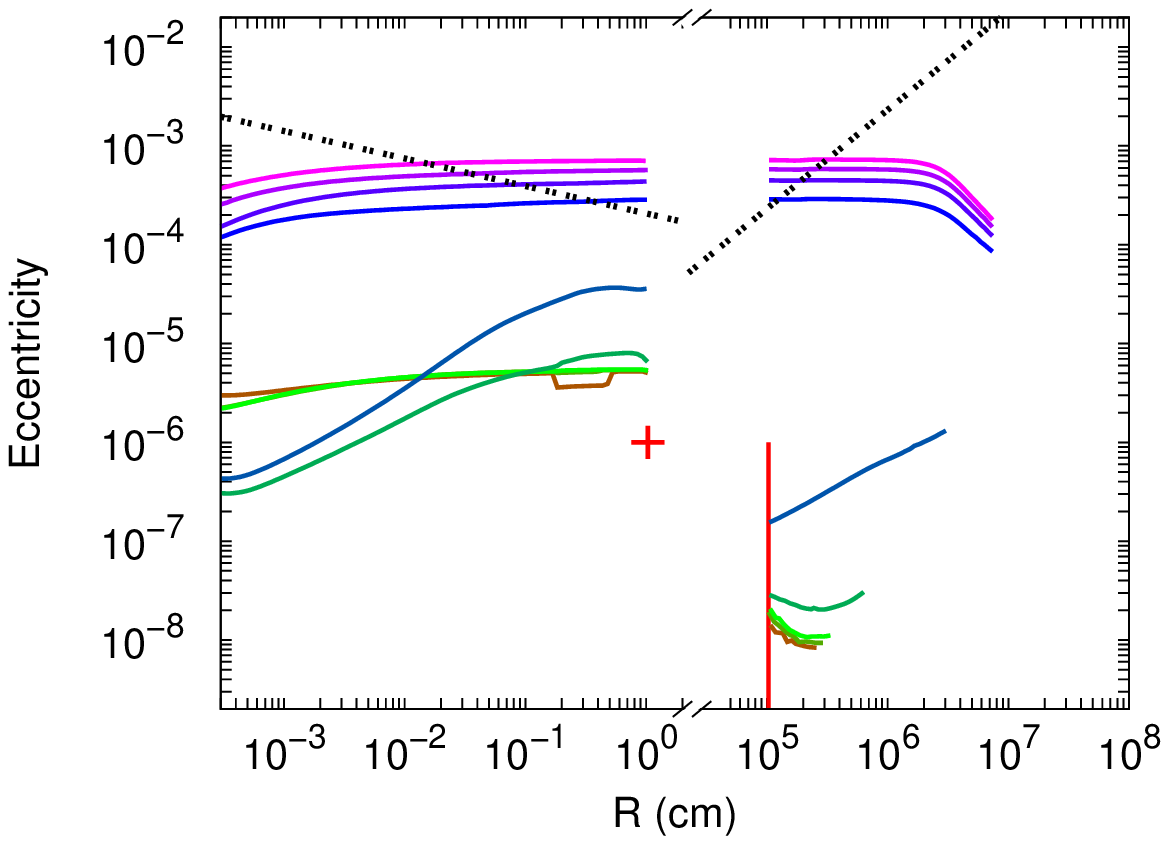}}
  \caption{Collisional growth, including the effects of cratering.
    Different curves are snapshots at $0$, $0.1, 0.2, 0.4, 0.8, 1.6,
    3.2, 6.4, 12.8$ and $25.6$ Myrs.  Growth of large bodies outpaces
    that in Fig. \ref{fig:plot318} in the early stage but is stalled
    after $\sim 2$ Myrs, when the largest bodies reach $\sim
    50\km$. The stalling occurs because the cm-grains are gradually
    down-sized and removed from the vicinity. The efficiency at this
    point is $\sim 5\%$.  The right-hand panel shows the eccentricity
    evolution, in particular, the dashed curves correspond to our
    adopted threshold for catastrophic destruction by an equal-mass
    projectile (eq. \ref{eq:Qstar} reduced by a factor of 4).
    Cratering is important even well below this threshold {\z due to
      the presence of smaller projectiles. This is increasingly
      important in late times when viscous stirring increases the
      dynamical excitation.}
}
\label{fig:cratering}
\end{figure*}

The results are displayed in Fig. \ref{fig:cratering}.  Cratering affects the outcome. Small bodies are now quickly
collisionally diminuated. Within a Myr, the cm-grains have largely been
removed. Growth of large bodies is initially faster but then is
truncated at $R \sim 50\km$. At this point, the mass in the big bodies
is $\sim 5\%$ of the total mass.  This value is similar to the case without cratering, even
though the largest bodies have only reached a fraction of the size of
that case. This is because the small body size, $s$, is reduced over
time in the cratering simulation (see eq. \ref{eq:yssigma}). The form
of the size spectrum is similar to that in the fiducial case.

In summary, cratering (and the subsequent blow-out) removes most of
the small grains quickly,
and arrests the collisional growth at an early stage.
 It also reduces the overall efficiency of large body
formation, to $\sim 5\%$ in our specific case.

\section{Discussion}
\label{sec:size}

Our work is motivated by the presence of the Cold Classical Kuiper
belt, which is hard to explain in the framework of collisionless
conglomeration. Here, we compare our simulation results against
observations of these bodies. We then proceed to discuss parameter
dependence of our models, as well as an alternative formation model,
the streaming instability.

\subsection{Matching the cold classical Kuiper belt -- Large KBOs}
\label{subsec:large}

\begin{figure}
  \centering
  {\includegraphics[width=0.49\textwidth,trim = 0 0 0 0, clip]{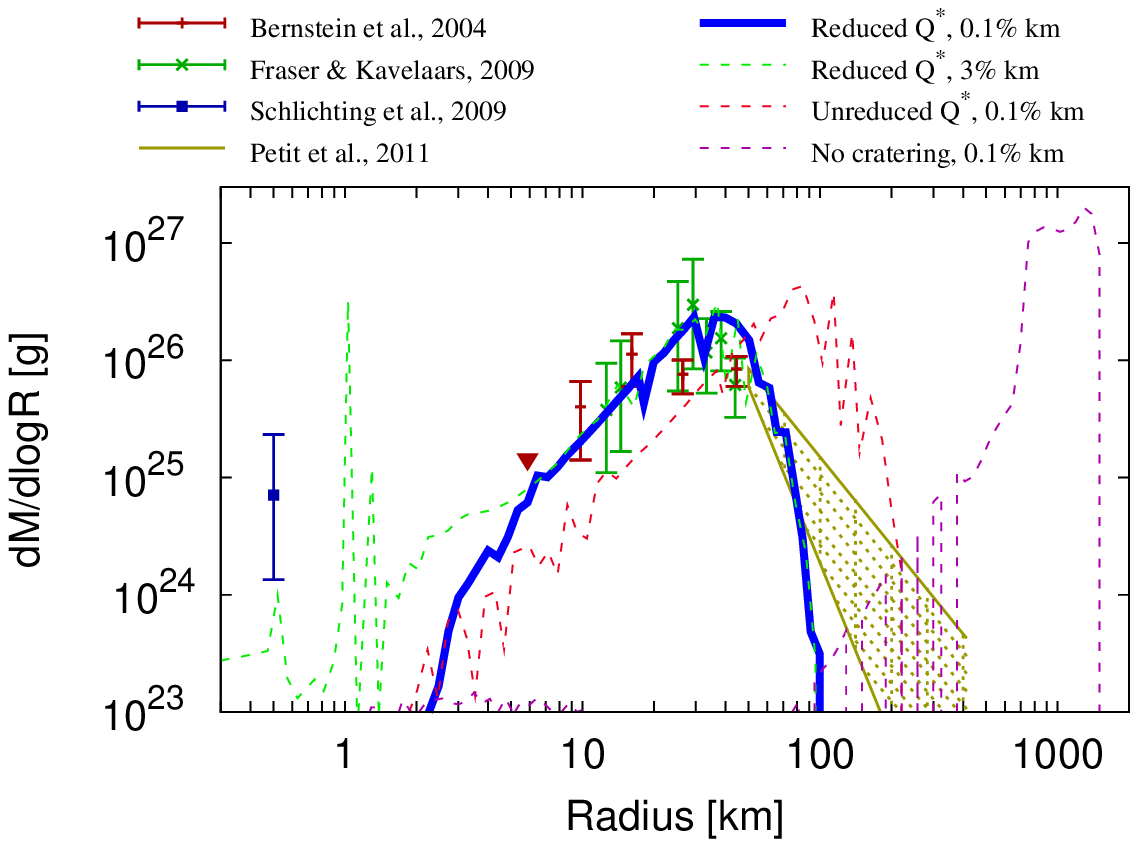}}
  \caption{Simulated and observed differential mass spectrum for Cold
    Classical objects. The observed mass distribution of Cold Classicals
    peak around $\sim 30\km$, and fall off on both sides.  
See text for details of conversion.
All four simulations here posit an initial surface density of $\sigma
= 3 \sigma_{\rm MMKB}$ ($0.3 M_\oplus$ within $42$ to $48$ AU).  The
cratering simulation in Fig. \ref{fig:cratering}, where we reduce the
material strength by a factor of $4$ from eq. \ref{eq:Qstar}, and
deposit $0.1\%$ of the initial mass in km-sized bodies, is shown as
solid blue line. Two other cratering simulations, one with more
initial mass in the km-bodies ($3\%$), the other with unreduced
strength, are shown as dotted lines, together with the result from the
no-cratering simulation (Fig. \ref{fig:plot318}). The observed mass
peak at $\sim 30\km$ is replicated in the cratering runs that have
reduced strength, while runs with no cratering, or with cratering but
at unreduced strength produce bodies that are too large by comparison.
Varying the initial mass fraction in km-bodies do not impact the
results. 
}
\label{fig:observation}
\end{figure}

We first synthesize various surveys to obtain the observed size
distribution of large bodies in the Cold Classical belt.  We
focus on surveys that either distill Cold Classicals from the
other populations using dynamical information, or ones that are only
sensitive to low inclination objects. The overlapping nature of the
trans-Neptunian populations means that pollution by resonant or
scattered populations that happen to lie at low inclinations remains a
possibility. For ease of comparison, we convert all observational
constraints to differential mass distribution, following
\citet{2009AJ....137...72F} and adopting an albedo of $6\%$. For
pencil-beam surveys \citep{Bernstein, 2009Natur.462..895S,
  2010Icar..210..944F}, we convert the observed number densities to
the total surface densities by assuming a scale height of $5
\deg$.

The relevant surveys are:
\citet{2011AJ....142..131P}, targeting objects with
cold-classical-like orbits with sizes down to $\sim 50\km$; 
\citet{Bernstein}, a pencil-beam HST survey restricting to objects
with $i < 5 \deg$ and sensitive down
to $\sim 10 \km$; 
\citet{2010Icar..210..944F}, similar in selection and sensitivity;
and lastly, the serendipitous discovery of a single $R \sim 0.5\km$
object by \citet{2009Natur.462..895S} using stellar occulation -- the
dynamical class of this object is not well constrained except that it
places an upper limit to the Cold Classicals at this size range.  A
somewhat weaker constraint is published by the TAOs survey
\citep{2010AJ....139.1499B}.  Combined together, these surveys
indicate that the largest Cold Classicals have radius $\sim 200$~km,
with the mass peaking around $30\km$. The differential mass $\left( dM/dlogR \right)$~very roughly falls off as $\sim R^{-2}$~above $30\km$ \citep{2011AJ....142..131P}, and as $\sim R^{1}$
(or possibly steeper) below $30 \km$~(see
  Fig. \ref{fig:observation}).

Collisional growth without cratering (Fig. \ref{fig:plot318}) produce
bodies that are too large compared to the observed belt. Cratering
simulations, on the other hand, can be easily adapted to reproduce the
observed size distribution, as is shown in
Fig. \ref{fig:observation}. With a material strength that is reduced
from the \citep{2009ApJ...691L.133S} value by a factor of 4, well
within the range of uncertainty, we are able to halt the
conglomeration growth and reproduce the observed mass peak at $\sim
30\km$.  Starting with an initial belt of $0.3 M_\oplus$, our simple
model reproduces the overall size distribution of Cold Classicals,
with a possible exception at very large sizes ($R \geq 100\km$).

While cratering is one possible way to arrest growth at intermediate
size, it is possible that external heating by a now-absent perturber (or by a more eccentric Neptune) can also do the job. This has
the advantage of explaining the current dynamical excitation of the
Cold Classicals ($e\sim 0.05$, higher than self-excitation should have
produced).

\subsection{A collisional break at $30\km$?}
\label{subsec:break}

The mass fall-off (also called `roll-over', `power-law break')
observed at below $\sim 30 \km$
\citep{Bernstein,2008AJ....136...83F,2009AJ....137...72F} arises
naturally in our growth model. However, in the past, this has instead
been interpreted as a collisional break, i.e., bodies below this size
have been removed by collisional destructions
\citep{PanSari,Kenyon:2004,Benavidez,FraserDivot}.\footnote{But also
  see \citealp{MorbidelliTrojan} who attribute it to an observational
  bias.}  However, this interpretation relies on an initially flat
mass distribution, $dM/d\log R \propto R^{0}$ (or a size distribution
$dN/d R \propto R^{-4}$) to have sufficient small bullets to break up
the large bodies. Such a mass distribution, however, obtained from
collisionless growth of a very massive disk, is astrophysically
unviable (see \S \ref{subsec:exclude}).

Intriguingly, a roll-over at roughly the same size has also been
observed for Jupiter Trojans \citep{2000AJ....120.1140J,
  Szabo,2008PASJ...60..297Y} and Neptune Trojans
\citep{SheppardTrujillo}. Their differential mass distributions can
all be roughly summarized as $dM/d\log R \propto R^{2}$ at sizes below
the peak and $dM/d\log R \propto R^{-2}$ above the peak, a shape also
found in the Scattered disk \citep{2012A&A...541A..94V}, but with a
larger roll-over size ($\sim$ 300 km).  It is difficult to ascribe
these features all as collisional breaks, as the physical environments
and therefore the collision frequencies are drastically different
among these objects. In contrast, a formation model of collisional
growth naturally accounts for such a peak.  We note that
\citet{Weidenschilling11} has proposed a formation model similar to
ours for the asteroid belt, which also exhibits a break at $\sim
50\km$.

Integrating the size distribution (e.g., solid line
in Fig. \ref{fig:observation}) forward for $4.5$ Gyrs, we find
that it suffers little collisional erosion. The expected fractional
dust luminosity at current time is $\sim 10^{-8}$, broadly consistent
with the upper and lower limits set by \citet{Teplitzetal:1999}
and \citet{Landgraf:2002}.

\subsection{Parameter Dependencies}
\label{subsec:dependence}

We briefly discuss how various model parameters affect the simulation
outcome. Much of this discussion follows the analytical scalings in
\S\ref{sec:cat}.

\begin{itemize}

\item If we scale up the initial surface density, larger bodies can form. 

\item If the size of the small grains, $s$, is reduced, we expect a higher
formation efficiency faster. 

\item In contrast, the size
of the large seeds, $1\km$ in our case, does not much influence the
outcome. 

\item Similarly, the initial mass fraction in these large seeds makes
  little difference to the final size spectrum, as is seen in
  Fig. \ref{fig:observation}, as long as these seeds remain a minor
  component.

\item If these seeds are totally absent, however, the initial growth
  will have to rely on the uncertain process of particle sticking, and
  a different growth scenario may occur
  \citep{2014ApJ...780...22L,2015ApJ...806...42K}.

\item If the material strength, relevant for the outcome of cratering
  collisions, is decreased, small bodies can be more easily destroyed,
  leading to a shortened growth episode and a smaller maximum size
  (Fig. \ref{fig:observation}).  {\z Conversely, if material strength
    is stronger than assumed here, the maximum size increases. This
    would be necessary if the mass peak is around $70\km$ as found by
    \citet{2014ApJ...782..100F}.}

\end{itemize}

\subsection{Alternative Formation Theories}

Is it possible that large Kuiper belt objects form via other routes,
as opposed to the gradual conglomeration as investigated here?
Alternatives include the direct collapse of a self-gravitating disk
\citep{Toomre,Goldreichward}; {\y incremental growth by direct
  sticking of small grains
  \citep[e.g.][]{1980Icar...44..172W,2005A&A...434..971D,2012A&A...540A..73W};}
the streaming instability in a gas-solid coupled fluid
\citep{youdingoodman,Johansen:2007}; abnormal concentration of dust
grains by turbulence \citep{2010Icar..208..518C}; and the accretion of
`pebbles' (cm-sized particles) by larger planetesimals in a gas-rich
disk \citep{pebble}.

The first option, the Toomre instability, could be important if the
disk initially contained only small grains. {\y If the grains are able
  to avoid stirring by gas turbulence of the type argued by
  \citet{1995Icar..116..433W} and continue} to cool below $e=10^{-6}$
to  reach Toomre number $Q = 1$, preliminary calculations \citep{michikoshi07,michikoshi09}
indicate that some Toomre mass scale bodies may form.  These are of
  order km in size for our parameters.  This then returns us to
exactly the scenario we posit here. So the Toomre instability may
provide the backdrop for our work.

In a gaseous nebula, direct sticking of small grains may not be efficient in producing ever larger bodies, even though naive
  calculations (perfect sticking) predict otherwise. This is because
  as bodies grow, collisions may become increasingly destructive and
  be dominated by fragmentation \citep{2005A&A...434..971D,Brauer}. 

  What about the popular streaming instability?  If the primordial
  Kuiper belt, when gas was present, was depleted in solid materials
  relative to that of MMSN, the streaming instability may be difficult
  to initiate. If it was of MMSN abundance, on the other hand, then it
  needs to convert only $\sim 0.1\%$ of the initial solid into large
  bodies. {\z It is not clear if this is a natural outcome of the streaming instability --
    although to be fair, our mechanism here started from an initial
    condition that we have also not yet justified. We note that there
    are recent attempts to predict the size distribution resulting
    from the streaming instability
    \citep{2015SciA....115109J,2015arXiv151200009S} and more work is
    needed to quantitatively compare these against the observed Cold
    Classical objects.}


\citet{2010Icar..208..518C} has proposed that small grains in a
  turbulent gas disk may be concentrated at certain regions to such a
  high density as to be self-gravitationally unstable. This proposal
  requires more physical understanding of the turbulence than is
  currently available.

`Pebble accretion' in a gas disk, on the other hand, shares some
similar attributes to the model presented here. While accretion in our
case is aided by the low dynamical excitation of cm-sized bodies as a
result of frequent collisions, accretion in that scenario is aided by
gas drag on cm-sized bodies. The size distribution arising from
`pebble accretion' is not yet investigated and is an interesting
future direction.

In summary, alternative theories are available but are not yet
investigated in sufficient detail to be tested.

\section{Summary}
\label{sec:discussion}

The consensus is building that the Cold Classical Kuiper Belt objects
likely formed {\it in situ} and have not suffered much dynamical
turmoil from the giant planets. They are reliable fossil records for
deciphering the environment and the mode of their formation.

It is often thought that the current-day anemic Kuiper belt is but a
skeletal remain of the primordial massive disk, given that models of
collisionless conglomeration requires a massive disk to form large
KBOs quickly and in sufficient numbers.  Moreover, it has been argued
that the observed `roll-over' at size $R \sim 30 \km$ is a result of
collisional diminuation over the past few billion years.  In this
study, we challenge both these ideas.  We argue that the Cold
Classicals may have formed with high efficiency, out of a low-mass
disk that is only a few percent of the MMSN value, and that the size
break at $30\km$ is primordial and may not be related to collisional
erosion.

In our model, much of the primordial mass is in small grains that
frequently collide and cool. This allows large KBOs to form quickly in
a low-density environment. The efficiency of large-body formation can
reach order unity, if growth continues undisturbed. 
The resultant mass distribution of large KBOs 
is top-heavy with $dM/d\log R \propto R^2$, in contrast with the flat
mass distribution from collisionless growth.  

Even in this low mass disk, conglomeration can, within a few million
years, convert almost all mass into bodies of size $\sim 1000\km$.  To
reproduce the observed peak at $\sim 30\km$, we introduce cratering
collisions. We show that, for plausible material strengths,
growth of large bodies can be arrested at an early stage, as small
grains can be demolished and removed from the system. Alternatively,
growth could also be arrested by, e.g., external excitation by nearby
planets \citep[e.g.,][]{1999Natur.402..635T,2005Natur.435..459T}, or a
stellar flyby in the denser birth environment
\citep{2005Icar..177..246K,2005Icar..173..559M}.

Processes not considered in our simple model include gas damping
\citep{Rafikov2003b}, gas dynamical friction
\citep{2015arXiv150302668G}, semi-collisional accretion
\citep{2007ApJ...658..593S}, and the influence of KBO binarity on
accretion.  Each could significantly affect the outcome of our model.

In conclusion, we suggest that Cold Classical KBOs likely formed not
out of a MMSN of solid material, but out of a `minimum mass Kuiper
belt' (MMKB), one that is but $\sim 1\%$~of the MMSN surface
density. This belt, with its low density, may well represent the true
outer edge of the Solar System. This edge could explain why Neptune's
outward migration is stalled at 30 au, and we might therefore expect no more large
bodies to be formed beyond the Cold Classical region.

\acknowledgements

We thank Eugene Chiang, Alexander Krivov, Chris Ormel and Hilke
Schlichting for helpful discussions. {\z We are also grateful for an
  anonymous referee who compels us to examine our assumptions
  closely.}  YW acknowledges grants from NSERC.  YL acknowledges
grants AST-1109776 and AST-1352369 from NSF, and NNX14AD21G from NASA.
AS was supported by the government of Ontario by a Ontario Graduate
Scholarship in Science and Technology; and is supported by the
European Union through ERC grant number 279973.

\bibliographystyle{apj}
\bibliography{kbogrowth}

\begin{thebibliography}{103}
\expandafter\ifx\csname natexlab\endcsname\relax\def\natexlab#1{#1}\fi

\bibitem[{{Barucci} {et~al.}(2008){Barucci}, {Boehnhardt}, {Cruikshank},
  {Morbidelli}, \& {Dotson}}]{2008ssbn.book.....B}
{Barucci}, M.~A., {Boehnhardt}, H., {Cruikshank}, D.~P., {Morbidelli}, A., \&
  {Dotson}, R. 2008, {The Solar System Beyond Neptune}

\bibitem[{{Benavidez} \& {Campo Bagatin}(2009)}]{Benavidez}
{Benavidez}, P.~G. \& {Campo Bagatin}, A. 2009, \planss, 57, 201

\bibitem[{{Bernstein} {et~al.}(2004){Bernstein}, {Trilling}, {Allen}, {Brown},
  {Holman}, \& {Malhotra}}]{Bernstein}
{Bernstein}, G.~M., {Trilling}, D.~E., {Allen}, R.~L., {Brown}, M.~E.,
  {Holman}, M., \& {Malhotra}, R. 2004, \aj, 128, 1364

\bibitem[{{Bianco} {et~al.}(2010){Bianco}, {Zhang}, {Lehner}, {Mondal}, {King},
  {Giammarco}, {Holman}, {Coehlo}, {Wang}, {Alcock}, {Axelrod}, {Byun}, {Chen},
  {Cook}, {Dave}, {de Pater}, {Kim}, {Lee}, {Lin}, {Lissauer}, {Marshall},
  {Protopapas}, {Rice}, {Schwamb}, {Wang}, \& {Wen}}]{2010AJ....139.1499B}
{Bianco}, F.~B., {Zhang}, Z., {Lehner}, M.~J., {Mondal}, S., {King}, S.,
  {Giammarco}, J., {Holman}, M.~J., {Coehlo}, N.~K., {Wang}, J., {Alcock}, C.,
  {Axelrod}, T., {Byun}, Y., {Chen}, W.~P., {Cook}, K.~H., {Dave}, R., {de
  Pater}, I., {Kim}, D., {Lee}, T., {Lin}, H., {Lissauer}, J.~J., {Marshall},
  S.~L., {Protopapas}, P., {Rice}, J.~A., {Schwamb}, M.~E., {Wang}, S., \&
  {Wen}, C. 2010, \aj, 139, 1499

\bibitem[{{Blum} \& {Wurm}(2008)}]{BlumWurm}
{Blum}, J. \& {Wurm}, G. 2008, \araa, 46, 21

\bibitem[{{Brauer} {et~al.}(2008){Brauer}, {Dullemond}, \& {Henning}}]{Brauer}
{Brauer}, F., {Dullemond}, C.~P., \& {Henning}, T. 2008, \aap, 480, 859

\bibitem[{{Brown}(2001)}]{2001AJ....121.2804B}
{Brown}, M.~E. 2001, \aj, 121, 2804

\bibitem[{{Brucker} {et~al.}(2009){Brucker}, {Grundy}, {Stansberry}, {Spencer},
  {Sheppard}, {Chiang}, \& {Buie}}]{2009Icar..201..284B}
{Brucker}, M.~J., {Grundy}, W.~M., {Stansberry}, J.~A., {Spencer}, J.~R.,
  {Sheppard}, S.~S., {Chiang}, E.~I., \& {Buie}, M.~W. 2009, \icarus, 201, 284

\bibitem[{{Bryden} {et~al.}(2006){Bryden}, {Beichman}, {Trilling}, {Rieke},
  {Holmes}, {Lawler}, {Stapelfeldt}, {Werner}, {Gautier}, {Blaylock}, {Gordon},
  {Stansberry}, \& {Su}}]{2006ApJ...636.1098B}
{Bryden}, G., {Beichman}, C.~A., {Trilling}, D.~E., {Rieke}, G.~H., {Holmes},
  E.~K., {Lawler}, S.~M., {Stapelfeldt}, K.~R., {Werner}, M.~W., {Gautier},
  T.~N., {Blaylock}, M., {Gordon}, K.~D., {Stansberry}, J.~A., \& {Su},
  K.~Y.~L. 2006, \apj, 636, 1098

\bibitem[{{Cuzzi} {et~al.}(2010){Cuzzi}, {Hogan}, \&
  {Bottke}}]{2010Icar..208..518C}
{Cuzzi}, J.~N., {Hogan}, R.~C., \& {Bottke}, W.~F. 2010, \icarus, 208, 518

\bibitem[{{Dawson} \& {Murray-Clay}(2012)}]{2012ApJ...750...43D}
{Dawson}, R.~I. \& {Murray-Clay}, R. 2012, \apj, 750, 43

\bibitem[{{Dohnanyi}(1969)}]{Dohnanyi:1969}
{Dohnanyi}, J.~W. 1969, \jgr, 74, 2531

\bibitem[{{Dullemond} \& {Dominik}(2005)}]{2005A&A...434..971D}
{Dullemond}, C.~P. \& {Dominik}, C. 2005, \aap, 434, 971

\bibitem[{{Edgeworth}(1949)}]{1949MNRAS.109..600E}
{Edgeworth}, K.~E. 1949, \mnras, 109, 600

\bibitem[{{Elliot} {et~al.}(2005){Elliot}, {Kern}, {Clancy}, {Gulbis},
  {Millis}, {Buie}, {Wasserman}, {Chiang}, {Jordan}, {Trilling}, \&
  {Meech}}]{2005AJ....129.1117E}
{Elliot}, J.~L., {Kern}, S.~D., {Clancy}, K.~B., {Gulbis}, A.~A.~S., {Millis},
  R.~L., {Buie}, M.~W., {Wasserman}, L.~H., {Chiang}, E.~I., {Jordan}, A.~B.,
  {Trilling}, D.~E., \& {Meech}, K.~J. 2005, \aj, 129, 1117

\bibitem[{{Fernandez} \& {Ip}(1984)}]{1984Icar...58..109F}
{Fernandez}, J.~A. \& {Ip}, W.-H. 1984, \icarus, 58, 109

\bibitem[{{Fraser} \& {Brown}(2011)}]{2011AAS...21730604F}
{Fraser}, W. \& {Brown}, M.~E. 2011, in Bulletin of the American Astronomical
  Society, Vol.~43, American Astronomical Society Meeting Abstracts \#217,
  \#306.04--+

\bibitem[{{Fraser}(2009)}]{FraserDivot}
{Fraser}, W.~C. 2009, \apj, 706, 119

\bibitem[{{Fraser} {et~al.}(2014){Fraser}, {Brown}, {Morbidelli}, {Parker}, \&
  {Batygin}}]{2014ApJ...782..100F}
{Fraser}, W.~C., {Brown}, M.~E., {Morbidelli}, A., {Parker}, A., \& {Batygin},
  K. 2014, \apj, 782, 100

\bibitem[{{Fraser} {et~al.}(2010){Fraser}, {Brown}, \&
  {Schwamb}}]{2010Icar..210..944F}
{Fraser}, W.~C., {Brown}, M.~E., \& {Schwamb}, M.~E. 2010, \icarus, 210, 944

\bibitem[{{Fraser} \& {Kavelaars}(2009)}]{2009AJ....137...72F}
{Fraser}, W.~C. \& {Kavelaars}, J.~J. 2009, \aj, 137, 72

\bibitem[{{Fuentes} \& {Holman}(2008)}]{2008AJ....136...83F}
{Fuentes}, C.~I. \& {Holman}, M.~J. 2008, \aj, 136, 83

\bibitem[{{Goldreich} {et~al.}(2004){Goldreich}, {Lithwick}, \& {Sari}}]{GLS}
{Goldreich}, P., {Lithwick}, Y., \& {Sari}, R. 2004, \araa, 42, 549

\bibitem[{{Goldreich} \& {Ward}(1973)}]{Goldreichward}
{Goldreich}, P. \& {Ward}, W.~R. 1973, \apj, 183, 1051

\bibitem[{{Gomes} {et~al.}(2008){Gomes}, {Fern{\'a}ndez}, {Gallardo}, \&
  {Brunini}}]{2008ssbn.book..259G}
{Gomes}, R.~S., {Fern{\'a}ndez}, J.~A., {Gallardo}, T., \& {Brunini}, A. {The
  Scattered Disk: Origins, Dynamics, and End States}, ed. {Barucci, M.~A.,
  Boehnhardt, H., Cruikshank, D.~P., Morbidelli, A., \& Dotson, R.}, 259--273

\bibitem[{{Gomes} {et~al.}(2004){Gomes}, {Morbidelli}, \&
  {Levison}}]{2004Icar..170..492G}
{Gomes}, R.~S., {Morbidelli}, A., \& {Levison}, H.~F. 2004, \icarus, 170, 492

\bibitem[{{Greenberg} {et~al.}(1978){Greenberg}, {Hartmann}, {Chapman}, \&
  {Wacker}}]{1978Icar...35....1G}
{Greenberg}, R., {Hartmann}, W.~K., {Chapman}, C.~R., \& {Wacker}, J.~F. 1978,
  \icarus, 35, 1

\bibitem[{{Grishin} \& {Perets}(2015)}]{2015arXiv150302668G}
{Grishin}, E. \& {Perets}, H.~B. 2015, ArXiv e-prints

\bibitem[{{Hayashi}(1981)}]{Hayashi}
{Hayashi}, C. 1981, Progress of Theoretical Physics Supplement, 70, 35

\bibitem[{{Ida} \& {Makino}(1992)}]{1992Icar...96..107I}
{Ida}, S. \& {Makino}, J. 1992, \icarus, 96, 107

\bibitem[{{Isella} {et~al.}(2012){Isella}, {P{\'e}rez}, \&
  {Carpenter}}]{2012ApJ...747..136I}
{Isella}, A., {P{\'e}rez}, L.~M., \& {Carpenter}, J.~M. 2012, \apj, 747, 136

\bibitem[{{Jewitt} \& {Luu}(1993)}]{1993Natur.362..730J}
{Jewitt}, D. \& {Luu}, J. 1993, \nat, 362, 730

\bibitem[{{Jewitt} {et~al.}(2000){Jewitt}, {Trujillo}, \&
  {Luu}}]{2000AJ....120.1140J}
{Jewitt}, D.~C., {Trujillo}, C.~A., \& {Luu}, J.~X. 2000, \aj, 120, 1140

\bibitem[{{Johansen} {et~al.}(2015){Johansen}, {Mac Low}, {Lacerda}, \&
  {Bizzarro}}]{2015SciA....115109J}
{Johansen}, A., {Mac Low}, M.-M., {Lacerda}, P., \& {Bizzarro}, M. 2015,
  Science Advances, 1, 15109

\bibitem[{{Johansen} {et~al.}(2007){Johansen}, {Oishi}, {Low}, {Klahr},
  {Henning}, \& {Youdin}}]{Johansen:2007}
{Johansen}, A., {Oishi}, J.~S., {Low}, M.-M.~M., {Klahr}, H., {Henning}, T., \&
  {Youdin}, A. 2007, \nat, 448, 1022

\bibitem[{{Kelley} {et~al.}(2013){Kelley}, {Lindler}, {Bodewits}, {A'Hearn},
  {Lisse}, {Kolokolova}, {Kissel}, \& {Hermalyn}}]{2013Icar..222..634K}
{Kelley}, M.~S., {Lindler}, D.~J., {Bodewits}, D., {A'Hearn}, M.~F., {Lisse},
  C.~M., {Kolokolova}, L., {Kissel}, J., \& {Hermalyn}, B. 2013, \icarus, 222,
  634

\bibitem[{{Kenyon} \& {Bromley}(2004)}]{Kenyon:2004}
{Kenyon}, S.~J. \& {Bromley}, B.~C. 2004, \aj, 128, 1916

\bibitem[{{Kenyon} \& {Bromley}(2008)}]{2008ApJS..179..451K}
---. 2008, \apjs, 179, 451

\bibitem[{{Kenyon} \& {Bromley}(2015)}]{2015ApJ...806...42K}
---. 2015, \apj, 806, 42

\bibitem[{{Kenyon} \& {Luu}(1998)}]{1998AJ....115.2136K}
{Kenyon}, S.~J. \& {Luu}, J.~X. 1998, \aj, 115, 2136

\bibitem[{{Kenyon} \& {Luu}(1999)}]{1999ApJ...526..465K}
---. 1999, \apj, 526, 465

\bibitem[{{Kobayashi} {et~al.}(2005){Kobayashi}, {Ida}, \&
  {Tanaka}}]{2005Icar..177..246K}
{Kobayashi}, H., {Ida}, S., \& {Tanaka}, H. 2005, \icarus, 177, 246

\bibitem[{{Kobayashi} \& {Tanaka}(2010)}]{2010Icar..206..735K}
{Kobayashi}, H. \& {Tanaka}, H. 2010, \icarus, 206, 735

\bibitem[{{Kokubo} \& {Ida}(1998)}]{1998Icar..131..171K}
{Kokubo}, E. \& {Ida}, S. 1998, \icarus, 131, 171

\bibitem[{{Kretke} \& {Levison}(2015)}]{2015Icar..262....9K}
{Kretke}, K.~A. \& {Levison}, H.~F. 2015, \icarus, 262, 9

\bibitem[{{Lacerda} {et~al.}(2014){Lacerda}, {Fornasier}, {Lellouch}, {Kiss},
  {Vilenius}, {Santos-Sanz}, {Rengel}, {M{\"u}ller}, {Stansberry}, {Duffard},
  {Delsanti}, \& {Guilbert-Lepoutre}}]{2014ApJ...793L...2L}
{Lacerda}, P., {Fornasier}, S., {Lellouch}, E., {Kiss}, C., {Vilenius}, E.,
  {Santos-Sanz}, P., {Rengel}, M., {M{\"u}ller}, T., {Stansberry}, J.,
  {Duffard}, R., {Delsanti}, A., \& {Guilbert-Lepoutre}, A. 2014, \apjl, 793,
  L2

\bibitem[{{Lambrechts} \& {Johansen}(2012)}]{pebble}
{Lambrechts}, M. \& {Johansen}, A. 2012, \aap, 544, A32

\bibitem[{{Landgraf} {et~al.}(2002){Landgraf}, {Liou}, {Zook}, \&
  {Gr{\"u}n}}]{Landgraf:2002}
{Landgraf}, M., {Liou}, J.-C., {Zook}, H.~A., \& {Gr{\"u}n}, E. 2002, \aj, 123,
  2857

\bibitem[{{Levison} {et~al.}(2008){Levison}, {Morbidelli}, {Vanlaerhoven},
  {Gomes}, \& {Tsiganis}}]{2008Icar..196..258L}
{Levison}, H.~F., {Morbidelli}, A., {Vanlaerhoven}, C., {Gomes}, R., \&
  {Tsiganis}, K. 2008, \icarus, 196, 258

\bibitem[{{Levison} \& {Stern}(2001)}]{2001AJ....121.1730L}
{Levison}, H.~F. \& {Stern}, S.~A. 2001, \aj, 121, 1730

\bibitem[{{Lissauer}(1987)}]{1987Icar...69..249L}
{Lissauer}, J.~J. 1987, \icarus, 69, 249

\bibitem[{{Lithwick}(2014)}]{2014ApJ...780...22L}
{Lithwick}, Y. 2014, \apj, 780, 22

\bibitem[{{Luu} {et~al.}(1997){Luu}, {Marsden}, {Jewitt}, {Trujillo},
  {Hergenrother}, {Chen}, \& {Offutt}}]{1997Natur.387..573L}
{Luu}, J., {Marsden}, B.~G., {Jewitt}, D., {Trujillo}, C.~A., {Hergenrother},
  C.~W., {Chen}, J., \& {Offutt}, W.~B. 1997, \nat, 387, 573

\bibitem[{{Malhotra}(1993)}]{1993Natur.365..819M}
{Malhotra}, R. 1993, \nat, 365, 819

\bibitem[{{Malhotra}(1995)}]{1995AJ....110..420M}
---. 1995, \aj, 110, 420

\bibitem[{{Massironi} \& {Simioni}(2015)}]{Massironi}
{Massironi}, M. \& {Simioni}, E. and.~{Marzari}, F. a. C. G. a. L. a. P. M. a.
  J. L. a. N. G. a. L. S. a. E. M. R. a. P. F. a. S.~F. 2015, Nature

\bibitem[{{Melita} {et~al.}(2005){Melita}, {Larwood}, \&
  {Williams}}]{2005Icar..173..559M}
{Melita}, M.~D., {Larwood}, J.~D., \& {Williams}, I.~P. 2005, \icarus, 173, 559

\bibitem[{{Michikoshi} {et~al.}(2007){Michikoshi}, {Inutsuka}, {Kokubo}, \&
  {Furuya}}]{michikoshi07}
{Michikoshi}, S., {Inutsuka}, S.-i., {Kokubo}, E., \& {Furuya}, I. 2007, \apj,
  657, 521

\bibitem[{{Michikoshi} {et~al.}(2009){Michikoshi}, {Kokubo}, \&
  {Inutsuka}}]{michikoshi09}
{Michikoshi}, S., {Kokubo}, E., \& {Inutsuka}, S.-i. 2009, \apj, 703, 1363

\bibitem[{{Morbidelli} {et~al.}(2014){Morbidelli}, {Gaspar}, \&
  {Nesvorny}}]{2014Icar..232...81M}
{Morbidelli}, A., {Gaspar}, H.~S., \& {Nesvorny}, D. 2014, \icarus, 232, 81

\bibitem[{{Morbidelli} {et~al.}(2009){Morbidelli}, {Levison}, {Bottke},
  {Dones}, \& {Nesvorn{\'y}}}]{MorbidelliTrojan}
{Morbidelli}, A., {Levison}, H.~F., {Bottke}, W.~F., {Dones}, L., \&
  {Nesvorn{\'y}}, D. 2009, \icarus, 202, 310

\bibitem[{{Noll} {et~al.}(2008){Noll}, {Grundy}, {Stephens}, {Levison}, \&
  {Kern}}]{2008Icar..194..758N}
{Noll}, K.~S., {Grundy}, W.~M., {Stephens}, D.~C., {Levison}, H.~F., \& {Kern},
  S.~D. 2008, \icarus, 194, 758

\bibitem[{{Ormel} {et~al.}(2010){Ormel}, {Dullemond}, \&
  {Spaans}}]{2010Icar..210..507O}
{Ormel}, C.~W., {Dullemond}, C.~P., \& {Spaans}, M. 2010, \icarus, 210, 507

\bibitem[{{Ormel} \& {Klahr}(2010)}]{ormel}
{Ormel}, C.~W. \& {Klahr}, H.~H. 2010, \aap, 520, A43

\bibitem[{{Pan} \& {Sari}(2005)}]{PanSari}
{Pan}, M. \& {Sari}, R. 2005, Icarus, 173, 342

\bibitem[{{Parker} \& {Kavelaars}(2010)}]{2010ApJ...722L.204P}
{Parker}, A.~H. \& {Kavelaars}, J.~J. 2010, \apjl, 722, L204

\bibitem[{{Parker} \& {Kavelaars}(2012)}]{2012ApJ...744..139P}
---. 2012, \apj, 744, 139

\bibitem[{{Parker} {et~al.}(2011){Parker}, {Kavelaars}, {Petit}, {Jones},
  {Gladman}, \& {Parker}}]{2011ApJ...743....1P}
{Parker}, A.~H., {Kavelaars}, J.~J., {Petit}, J.-M., {Jones}, L., {Gladman},
  B., \& {Parker}, J. 2011, \apj, 743, 1

\bibitem[{{P{\'e}rez} {et~al.}(2012){P{\'e}rez}, {Carpenter}, {Chandler},
  {Isella}, {Andrews}, {Ricci}, {Calvet}, {Corder}, {Deller}, {Dullemond},
  {Greaves}, {Harris}, {Henning}, {Kwon}, {Lazio}, {Linz}, {Mundy}, {Sargent},
  {Storm}, {Testi}, \& {Wilner}}]{2012ApJ...760L..17P}
{P{\'e}rez}, L.~M., {Carpenter}, J.~M., {Chandler}, C.~J., {Isella}, A.,
  {Andrews}, S.~M., {Ricci}, L., {Calvet}, N., {Corder}, S.~A., {Deller},
  A.~T., {Dullemond}, C.~P., {Greaves}, J.~S., {Harris}, R.~J., {Henning}, T.,
  {Kwon}, W., {Lazio}, J., {Linz}, H., {Mundy}, L.~G., {Sargent}, A.~I.,
  {Storm}, S., {Testi}, L., \& {Wilner}, D.~J. 2012, \apjl, 760, L17

\bibitem[{{Perna} {et~al.}(2010){Perna}, {Barucci}, {Fornasier}, {DeMeo},
  {Alvarez-Candal}, {Merlin}, {Dotto}, {Doressoundiram}, \& {de
  Bergh}}]{2010A&A...510A..53P}
{Perna}, D., {Barucci}, M.~A., {Fornasier}, S., {DeMeo}, F.~E.,
  {Alvarez-Candal}, A., {Merlin}, F., {Dotto}, E., {Doressoundiram}, A., \& {de
  Bergh}, C. 2010, \aap, 510, A53+

\bibitem[{{Petit} {et~al.}(2011){Petit}, {Kavelaars}, {Gladman}, {Jones},
  {Parker}, {Van Laerhoven}, {Nicholson}, {Mars}, {Rousselot}, {Mousis},
  {Marsden}, {Bieryla}, {Taylor}, {Ashby}, {Benavidez}, {Campo Bagatin}, \&
  {Bernabeu}}]{2011AJ....142..131P}
{Petit}, J.-M., {Kavelaars}, J.~J., {Gladman}, B.~J., {Jones}, R.~L., {Parker},
  J.~W., {Van Laerhoven}, C., {Nicholson}, P., {Mars}, G., {Rousselot}, P.,
  {Mousis}, O., {Marsden}, B., {Bieryla}, A., {Taylor}, M., {Ashby}, M.~L.~N.,
  {Benavidez}, P., {Campo Bagatin}, A., \& {Bernabeu}, G. 2011, \aj, 142, 131

\bibitem[{{Rafikov}(2003{\natexlab{a}})}]{Rafikov2003b}
{Rafikov}, R.~R. 2003{\natexlab{a}}, \aj, 126, 2529

\bibitem[{{Rafikov}(2003{\natexlab{b}})}]{Rafikov}
---. 2003{\natexlab{b}}, \aj, 125, 942

\bibitem[{{Safronov}(1969)}]{Safronov1969}
{Safronov}, V.~S. 1969, {Evoliutsiia doplanetnogo oblaka.}, ed. {Safronov,
  V.~S.}

\bibitem[{{Schlichting} {et~al.}(2009){Schlichting}, {Ofek}, {Wenz}, {Sari},
  {Gal-Yam}, {Livio}, {Nelan}, \& {Zucker}}]{2009Natur.462..895S}
{Schlichting}, H.~E., {Ofek}, E.~O., {Wenz}, M., {Sari}, R., {Gal-Yam}, A.,
  {Livio}, M., {Nelan}, E., \& {Zucker}, S. 2009, \nat, 462, 895

\bibitem[{{Schlichting} \& {Sari}(2007)}]{2007ApJ...658..593S}
{Schlichting}, H.~E. \& {Sari}, R. 2007, \apj, 658, 593

\bibitem[{{Schlichting} \& {Sari}(2011)}]{2011ApJ...728...68S}
---. 2011, \apj, 728, 68

\bibitem[{{Shannon} \& {Wu}(2011)}]{2011ApJ...739...36S}
{Shannon}, A. \& {Wu}, Y. 2011, \apj, 739, 36

\bibitem[{{Shannon} {et~al.}(2015){Shannon}, {Wu}, \&
  {Lithwick}}]{2015ApJ...801...15S}
{Shannon}, A., {Wu}, Y., \& {Lithwick}, Y. 2015, \apj, 801, 15

\bibitem[{{Sheppard} \& {Trujillo}(2010)}]{SheppardTrujillo}
{Sheppard}, S.~S. \& {Trujillo}, C.~A. 2010, \apjl, 723, L233

\bibitem[{{Simon} {et~al.}(2015){Simon}, {Armitage}, {Li}, \&
  {Youdin}}]{2015arXiv151200009S}
{Simon}, J.~B., {Armitage}, P.~J., {Li}, R., \& {Youdin}, A.~N. 2015, ArXiv
  e-prints

\bibitem[{{Stewart} \& {Leinhardt}(2009)}]{2009ApJ...691L.133S}
{Stewart}, S.~T. \& {Leinhardt}, Z.~M. 2009, \apjl, 691, L133

\bibitem[{{Szab{\'o}} {et~al.}(2007){Szab{\'o}}, {Ivezi{\'c}}, {Juri{\'c}}, \&
  {Lupton}}]{Szabo}
{Szab{\'o}}, G.~M., {Ivezi{\'c}}, {\v Z}., {Juri{\'c}}, M., \& {Lupton}, R.
  2007, \mnras, 377, 1393

\bibitem[{{Tegler} \& {Romanishin}(2000)}]{2000Natur.407..979T}
{Tegler}, S.~C. \& {Romanishin}, W. 2000, \nat, 407, 979

\bibitem[{{Teplitz} {et~al.}(1999){Teplitz}, {Stern}, {Anderson}, {Rosenbaum},
  {Scalise}, \& {Wentzler}}]{Teplitzetal:1999}
{Teplitz}, V.~L., {Stern}, S.~A., {Anderson}, J.~D., {Rosenbaum}, D.,
  {Scalise}, R.~J., \& {Wentzler}, P. 1999, \apj, 516, 425

\bibitem[{{Thommes} {et~al.}(1999){Thommes}, {Duncan}, \&
  {Levison}}]{1999Natur.402..635T}
{Thommes}, E.~W., {Duncan}, M.~J., \& {Levison}, H.~F. 1999, \nat, 402, 635

\bibitem[{{Toomre}(1964)}]{Toomre}
{Toomre}, A. 1964, \apj, 139, 1217

\bibitem[{{Trotta} {et~al.}(2013){Trotta}, {Testi}, {Natta}, {Isella}, \&
  {Ricci}}]{2013A&A...558A..64T}
{Trotta}, F., {Testi}, L., {Natta}, A., {Isella}, A., \& {Ricci}, L. 2013,
  \aap, 558, A64

\bibitem[{{Trujillo} \& {Brown}(2002)}]{TrujilloBrown}
{Trujillo}, C.~A. \& {Brown}, M.~E. 2002, \apjl, 566, L125

\bibitem[{{Tsiganis} {et~al.}(2005){Tsiganis}, {Gomes}, {Morbidelli}, \&
  {Levison}}]{2005Natur.435..459T}
{Tsiganis}, K., {Gomes}, R., {Morbidelli}, A., \& {Levison}, H.~F. 2005, \nat,
  435, 459

\bibitem[{{Vilenius} {et~al.}(2012){Vilenius}, {Kiss}, {Mommert}, {M{\"u}ller},
  {Santos-Sanz}, {Pal}, {Stansberry}, {Mueller}, {Peixinho}, {Fornasier},
  {Lellouch}, {Delsanti}, {Thirouin}, {Ortiz}, {Duffard}, {Perna}, {Szalai},
  {Protopapa}, {Henry}, {Hestroffer}, {Rengel}, {Dotto}, \&
  {Hartogh}}]{2012A&A...541A..94V}
{Vilenius}, E., {Kiss}, C., {Mommert}, M., {M{\"u}ller}, T., {Santos-Sanz}, P.,
  {Pal}, A., {Stansberry}, J., {Mueller}, M., {Peixinho}, N., {Fornasier}, S.,
  {Lellouch}, E., {Delsanti}, A., {Thirouin}, A., {Ortiz}, J.~L., {Duffard},
  R., {Perna}, D., {Szalai}, N., {Protopapa}, S., {Henry}, F., {Hestroffer},
  D., {Rengel}, M., {Dotto}, E., \& {Hartogh}, P. 2012, \aap, 541, A94

\bibitem[{{Vitense} {et~al.}(2010){Vitense}, {Krivov}, \&
  {L{\"o}hne}}]{2010A&A...520A..32V}
{Vitense}, C., {Krivov}, A.~V., \& {L{\"o}hne}, T. 2010, \aap, 520, A32+

\bibitem[{{Weidenschilling}(1977{\natexlab{a}})}]{1977MNRAS.180...57W}
{Weidenschilling}, S.~J. 1977{\natexlab{a}}, \mnras, 180, 57

\bibitem[{{Weidenschilling}(1977{\natexlab{b}})}]{1977Ap&SS..51..153W}
---. 1977{\natexlab{b}}, \apss, 51, 153

\bibitem[{{Weidenschilling}(1980)}]{1980Icar...44..172W}
---. 1980, Icarus, 44, 172

\bibitem[{{Weidenschilling}(1995)}]{1995Icar..116..433W}
---. 1995, \icarus, 116, 433

\bibitem[{{Weidenschilling}(2011)}]{Weidenschilling11}
---. 2011, \icarus, 214, 671

\bibitem[{{Weidenschilling} {et~al.}(1997){Weidenschilling}, {Spaute}, {Davis},
  {Marzari}, \& {Ohtsuki}}]{Weidenschilling}
{Weidenschilling}, S.~J., {Spaute}, D., {Davis}, D.~R., {Marzari}, F., \&
  {Ohtsuki}, K. 1997, \icarus, 128, 429

\bibitem[{{Wetherill} \& {Stewart}(1989)}]{WetherillStewart}
{Wetherill}, G.~W. \& {Stewart}, G.~R. 1989, \icarus, 77, 330

\bibitem[{{Windmark} {et~al.}(2012){Windmark}, {Birnstiel}, {G{\"u}ttler},
  {Blum}, {Dullemond}, \& {Henning}}]{2012A&A...540A..73W}
{Windmark}, F., {Birnstiel}, T., {G{\"u}ttler}, C., {Blum}, J., {Dullemond},
  C.~P., \& {Henning}, T. 2012, \aap, 540, A73

\bibitem[{{Yoshida} \& {Nakamura}(2008)}]{2008PASJ...60..297Y}
{Yoshida}, F. \& {Nakamura}, T. 2008, \pasj, 60, 297

\bibitem[{{Youdin} \& {Goodman}(2005)}]{youdingoodman}
{Youdin}, A.~N. \& {Goodman}, J. 2005, \apj, 620, 459

\bibitem[{{Zsom} {et~al.}(2010){Zsom}, {Ormel}, {G{\"u}ttler}, {Blum}, \&
  {Dullemond}}]{2010A&A...513A..57Z}
{Zsom}, A., {Ormel}, C.~W., {G{\"u}ttler}, C., {Blum}, J., \& {Dullemond},
  C.~P. 2010, \aap, 513, A57

\end{thebibliography}

\end{document}